\documentclass[twocolumn,twocolappendix]{aastex63}

\usepackage{color}
\usepackage{enumerate}
\usepackage{multirow}
\usepackage{graphicx}
\usepackage[normalem]{ulem}
\usepackage{xcolor}
\usepackage{hyperref}
\usepackage{amsmath}
\usepackage{kotex}

\def\Msun{{\rm M}_{\odot}}
\def\hMsun{\,h^{-1}{\rm M}_{\odot}}
\def\Mpc{\,h^{-1}{\rm {Mpc}}}
\def\kpc{\,h^{-1}{\rm {kpc}}}
\def\kms{\,{\rm km\,s^{-1}}}

\newcommand{\vect}[1]{\boldsymbol{#1}}

\revised{2019 November 22}
\accepted{2019 October 30}
\submitjournal{ApJ}

\shorttitle{Statistical Analysis of Flybys and Mergers of Dark Matter Halos}
\shortauthors{An et al.}

\begin{document}

\title{Living with Neighbors. II. Statistical Analysis of Flybys and Mergers of Dark Matter Halos in Cosmological Simulations}

\author[0000-0003-3791-0860]{Sung-Ho An}
\affiliation{Department of Astronomy, Yonsei University, Yonsei-ro 50, Seodaemun-gu, Seoul 03722, Republic of Korea}
\affiliation{Center for Galaxy Evolution Research, Yonsei University, Yonsei-ro 50, Seodaemun-gu, Seoul 03722, Republic of Korea}

\author[0000-0002-4391-2275]{Juhan Kim}
\affiliation{Center for Advanced Computation, Korea Institute for Advanced Study, Hoegiro 87, Dongdaemun-gu, Seoul 02455, Republic of Korea}

\author[0000-0001-7075-4156]{Jun-Sung Moon}
\affiliation{Department of Astronomy, Yonsei University, Yonsei-ro 50, Seodaemun-gu, Seoul 03722, Republic of Korea}
\affiliation{Center for Galaxy Evolution Research, Yonsei University, Yonsei-ro 50, Seodaemun-gu, Seoul 03722, Republic of Korea}

\author[0000-0002-1842-4325]{Suk-Jin Yoon}
\affiliation{Department of Astronomy, Yonsei University, Yonsei-ro 50, Seodaemun-gu, Seoul 03722, Republic of Korea}
\affiliation{Center for Galaxy Evolution Research, Yonsei University, Yonsei-ro 50, Seodaemun-gu, Seoul 03722, Republic of Korea}

\correspondingauthor{Suk-Jin Yoon}
\email{sjyoon0691@yonsei.ac.kr}

\begin{abstract}

We present a statistical analysis of flybys of dark matter halos compared to mergers using cosmological $N$-body simulations.
We mainly focus on gravitationally interacting target halos with mass of $10^{10.8}$\,--\,$10^{13.0}$\,$\hMsun$, and their neighbors are counted only when the mass ratio is 1:3\,--\,3:1 and the distance is less than the sum of the virial radii of target and neighbor.
The neighbors are divided into the flyby or merger samples if the pair's total energy is greater or smaller, respectively, than the capture criterion with consideration of dynamical friction.
The main results are as follows: 
(a) The flyby fraction increases by up to a factor of 50 with decreasing halo mass and by up to a factor of 400 with increasing large-scale density, while the merger fraction does not show any significant dependencies on these two parameters;
(b) The redshift evolution of the flyby fraction is twofold, increasing with redshift at 0\,$<$\,$z$\,$<$1 and remaining constant at $z$\,$>$\,1, while the merger fraction increases monotonically with redshift at $z=$ 0\,$\sim$\,4;
(c) The multiple interactions with two or more neighbors are on average flyby-dominated, and their fraction has a mass and environment dependence similar to that for the flyby fraction;
(d) Given that flybys substantially outnumber mergers toward $z=0$ (by a factor of five) and the multiple interactions are flyby-dominated, the flyby's contribution to galactic evolution is stronger than ever at the present epoch, especially for less massive halos and in the higher density environment.
We propose a scenario that connects the evolution of the flyby and merger fractions to the hierarchical structure formation process.

\end{abstract}

\keywords{Galaxy interactions (600), Galaxy encounters (592), Cosmological evolution (336), Galaxy environments (2029), Dark matter (353), Galaxy dark matter halos (1880), Large-scale structure of the universe (902), N-body simulations (1083)}

\section{Introduction}

The standard $\Lambda$CDM model of cosmology depicts that the present cosmic structures originate from tiny quantum fluctuations in the early universe.
Tiny, overdense regions are believed to gravitationally collapse and form virialized halos of dark matter.
According to the current paradigm of structure formation, a bigger halo forms later by merging smaller halos in a hierarchical manner \citep[e.g.,][]{Whi78a,Spr05}.
Therefore, numerous studies have acknowledged the merger as a key to understanding the cosmic evolution of galaxies \citep[e.g.,][]{Whi78b, Far82, Bar92, Bek95, Spr99, Cox06, Fak08, Lot08, Per14}.

Recently, however, the flyby encounter has emerged as a new and important type of galactic interaction with neighbors.
Although it has attracted less attention than the merger, the flyby may be a hidden driver behind dynamical and chemical processes that cannot be explained solely by a merger \citep[e.g.,][]{Wei06,Pat13}.
For instance, the existence of disk galaxies points to the importance of nonmerger processes because their disk structures are highly vulnerable to major mergers.
Moreover, the substructures in disk galaxies are not properly accounted for by mergers alone and invoke orbiting or passing neighbors.
\citet{Too72} and \citet{Ene73} showed that a close encounter by a galaxy in parabolic or hyperbolic orbit may be involved in the formation of tidal tails, spiral patterns, and galactic warps. 
Subsequent studies have further supported this hypothesis \citep[e.g.,][]{Tut06, Dub09, Don10, Kim14, Gom17}.
The tidal perturbation induced by flybying neighbors may create galactic bars \citep{Nog87, Miw98, Ber04, Lan14, Mar17}, ring structures \citep{You08}, and kinematically decoupled cores \citep{Hau94, Der04}.
Repeated flybys can expand galactic disks \citep{Lai14}, transform disk galaxies into lenticular galaxies \citep{Moo99, Gne03, Bek11, Vil14}, change the direction of the galaxy's angular momentum \citep{Bet12, Cen14}, and enhance \citep{Pat13} or quench \citep{Wet14} star formation activity.

There have been several efforts to quantify the importance of the flyby in a cosmological context. 
Early studies considered the galaxy cluster as a noteworthy place for the galactic interaction, especially for the flyby.
\citet{Gne03} found that the number of close encounters is even $\sim$1000 times higher than that of mergers for cluster galaxies.
\citet{Kne04} also investigated the flyby with other member galaxies or satellites in clusters and showed that about 30\,\% of cluster galaxies experience at least one flyby per orbit.
Recent cosmological $N$-body simulations also pointed out that a large fraction of halo pairs are unbound even as they are physically interacting.
\citet{Sin12} reported that contamination by flybys in counting bound pairs amounts to at least 20\,--\,30\,\% at $z$\,$<$\,3, which is similar to other measurements at $z$\,$\sim$\,0 \citep{Per06,Ton12}.
\citet{Mor13} also revealed that $\sim$62\% ($\sim$2\%) of satellite--satellite (central--satellite) pairs are unbound from each other.
A substantial fraction of galaxy (or halo) pairs ($\sim$50\%) identified at high redshifts ($z$\,$\gtrsim$\,1) do not merge until $z$\,=\,0 \citep{Jia12, Mor12, Sny17}.
Although the effect of the flyby has been consistently uncovered, it still remains unclear which parameters govern the frequency and properties of the flyby interaction.

This series of papers investigates both theoretically and observationally the characteristics of galactic interactions and their effect on the properties and evolution of galaxies in terms of dynamics and stellar populations.
\citet[][Paper I]{Moo19} showed the hydrodynamical effect of the nearest neighbor on star formation activity using the Sloan Digital Sky Survey \citep[SDSS;][]{Yor00}.
In the present second paper, we identify dark matter halos with their interacting neighbors in cosmological $N$-body simulations, statistically analyze the fractional contribution of flybys in comparison to mergers, and comparatively explore their evolution as functions of the halo mass, the large-scale environment, and the redshift.
This paper attempts to answer three main questions: 
(a) How frequently does a halo experience flyby events as functions of the halo mass, large-scale environment, and redshift, in comparison to mergers? 
(b) How strongly does the flyby affect the galactic evolution compared to the merger? 
(b) To what extent are the flyby fraction and the hierarchical structure formation related?

The paper is organized as follows. 
Section 2 describes the cosmological simulations and the halo-finding algorithm. 
Section 3 gives the definition of sampled halo pairs and their classification into flybys and mergers.
Section 4 comparatively examines the flyby interaction, the merger interaction, the total (flyby + merger) interaction, and the ``multiple'' interaction and reveals their behaviors depending on halo mass, environment, and redshift.
In Section 5, we discuss the physical causes of the dependence of the interaction fractions on the three parameters, and we link the flyby and merger fractions to the hierarchical structure formation.
Section 6 summarizes the results.

\section{Simulations and Halo Finding}

We use the Grid-of-Oct-Trees-Particle-Mesh code \citep[GOTPM;][]{Dub04}, which adopts a Tree-Particle-Mesh (TPM) scheme to measure the gravitational force on each particle under the periodic boundary conditions. GOTPM is fully parallelized using the MPI and OpenMP with variable-length domain decompositions of $Z$-directional slabs. GOTPM gets the initial power spectrum from the Code for Anisotropies in the Microwave Background \citep[CAMB;][]{Lew00} software.

We perform a set of 10 simulations using $512^3$ particles (GOTPM512) in a periodic cubic box with a side length of $L_{\rm box}$\,=\,$64\Mpc$.
The mass resolution of the simulations is $M_{\rm p}$\,=\,$1.55\times10^8\hMsun$.
To see the volume effect on the structure formation and to make Virgo-like clusters of halos, we additionally run another simulation with a bigger box of $L_{\rm box}$\,=\,$128\Mpc$ but with the same mass resolution using $1024^3$ particles (GOTPM1024).
For the entire 11 simulations, the fractions of matter and dark energy are $\Omega_m^0$\,=\,0.29 and $\Omega_{\Lambda}$\,=\,0.71, respectively, and the Hubble parameter at the current epoch is $H_0$\,=\,$100h\kms$\,${\rm Mpc}^{-1}$, where $h$\,=\,0.69 \citep[WMAP 9-year cosmology;][]{Ben13}. 
The softening length ($\epsilon$) is set to 1/10 of the mean displacement between particles, and $\epsilon$ is $12.5\kpc$ in comoving scale in all simulations.
We save 114 snapshots between $z$\,=\,129 and zero with a time spacing of $\sim$0.12 Gyr.

We utilize the Robust Overdensity Calculation using K-Space Topologically Adaptive Refinement \citep[ROCKSTAR;][]{Beh13}, which uses six-dimensional phase information (position and velocity) of halo particles.
\textsc{ROCKSTAR} applies the six-dimensional friend-of-friend (FoF) algorithm to identify member particles of (sub)halos.
To find virialized halos, the percolation linking length between particles is set to $l_{\rm FoF}$\,=\,0.28\,$d_{\rm mean}$, where $d_{\rm mean}$ is the mean particle separation. 
After finding the FoF halos, the mutual phase-space distance between member particles is used to divide the FoF halo into subhalos.
This phase-space scheme determines the particle membership of each subhalo by using the current circular maximum velocity ($v_{\rm max}$) and current velocity dispersion ($\sigma_{\rm v}$) of each subhalo, which is a major factor in stabilizing the particle membership.

Dark matter halo properties are defined based on the virial approximation from \citet{Bry98}. The virial mass ($M_{\rm vir}$) is a sum of masses of the member particles exceeding the virial density threshold ($\rho_{\rm vir}$). Then, the virial radius ($R_{\rm vir}$) is calculated with $R_{\rm vir}\equiv(3 M_{\rm vir}/4\pi\rho_{\rm vir})^{1/3}$.
For example, if we adopt $l_{\rm FoF}=0.28$\,$d_{\rm mean}$, a Milky Way-like halo with $M\sim 10^{12}\hMsun$ has a virial radius of $R_{\rm vir}\sim 200\kpc$ at $z=0$, which corresponds to the extent to which the mean overdensity of the halo is about 360 times the background density \citep[e.g.,][]{Kly11}.
At $z$\,=\,0, the number of halos extracted from the entire simulations is 2,046,538 with a halo mass greater than $10^{9.8}\hMsun$, which corresponds to the total mass of 40 particles.
Note that we do not distinguish central and satellite subhalos from our (sub)halo catalog, while \textsc{ROCKSTAR} identifies one central subhalo and satellite subhalos (if any) in each FoF halo.
From now on, a halo refers to a subhalo.

In identifying interacting halos, \textsc{ROCKSTAR} has two big advantages over other algorithms.
First, instead of using all member particles, it uses particles only in the innermost region (10\,\% of the halo radius) of a halo to define the position and velocity of the halo (see \citealt{Beh13} for details).
This reduces the effect of tidally induced, stretched structures that are due to the gravitational interaction and, therefore, allows the halo's position and velocity to trace the halo density peak well.
Second, \citet{Beh15} compared various halo-finding algorithms to examine their differences for major-merging halos at a given mass, and they found that phase-space finders recover well subhalo masses as well as positions and velocities from an input model. 
As the phase-space finder is able to trace the subhalos right before the coalescence, it derives a more accurate interaction fraction of identified halos.

Our simulations are sufficient for the statistical analysis of flyby and merger interactions for the following four reasons.
First, in this study we only consider halos with at least 100 member particles to suppress the noise in our statistical analysis of the interactions of halos.
\citet{Sin12} checked the mass-resolution effect on the number of interactions with a low- and high-resolution simulations of $512^{3}$ and $1024^{3}$ particles, respectively, in the same box size. 
They found that, if the analyzed halo consists of over 100 particles, the fluctuation in the number of halo interactions for a given mass scale does not exceed 1\,\%, implying that the mass resolution barely affects their conclusion.
Second, the effect of cosmic variance is reduced by running simulations with different random seed numbers used to generate the initial conditions.
This enhances the statistical significance of our data, and, accordingly, the maximum Poisson error ($\sqrt{N}/N$) substantially shrinks by 23.6\,\% when compared to one GOTPM512 simulation. 
Third, the GOTPM1024 simulation reproduces the larger-scale power better than the GOTPM512, and thus we can deal with the interaction fraction even for the Virgo-like clusters. 
In a comparison between 10 GOTPM512 simulations and one GOTPM1024 simulation, missing the large-scale power in the high-density environment does not significantly affect our conclusion (see Appendix \ref{sec:AA}).
Lastly, the time step ($\sim$0.12 Gyr) between two snapshots is smaller than the crossing time of Milky Way-sized halos, which is defined as $R/v$ where $R$ is a radius of the halo and $v$ is a rotation velocity with a circular orbit at $R$.
In the case of the $10^{12}\hMsun$ halo, the crossing time at $z$\,=\,0 is $\sim$1 Gyr (200\,kpc\,/\,$200\kms$), which corresponds to about eight steps over the course of tidal interactions.

\section{Analysis}

\begin{figure*}[ht!]
\centering
\includegraphics[width=\textwidth]{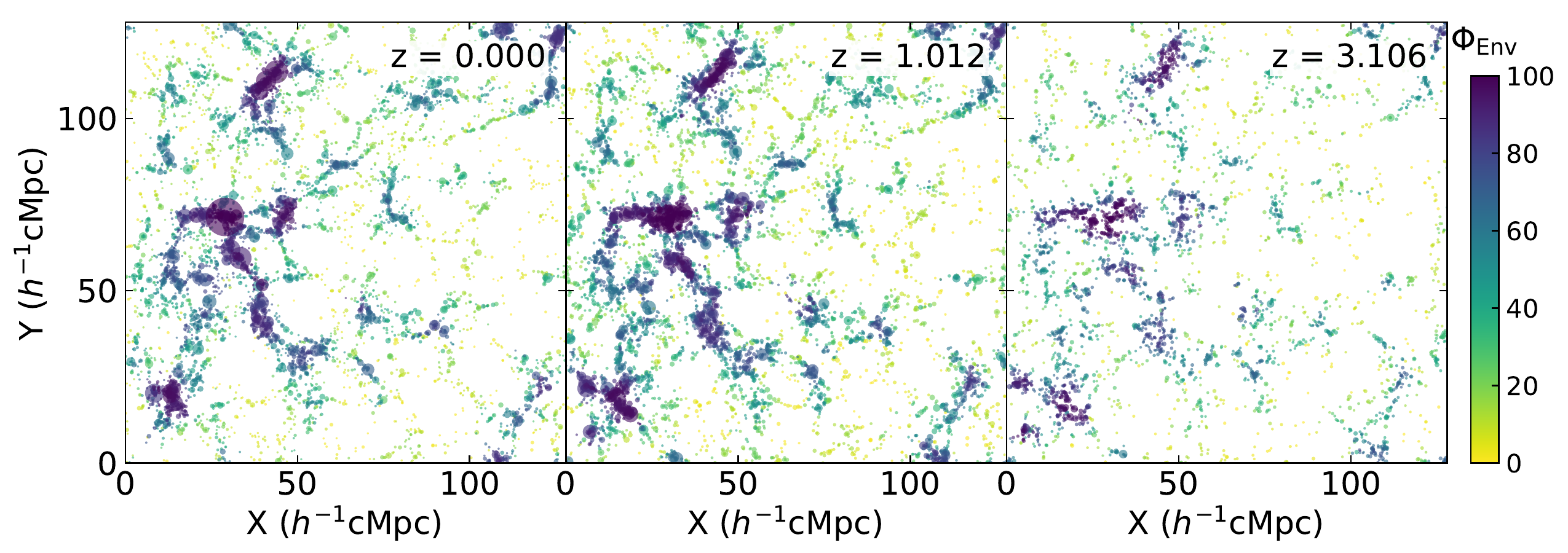}
\caption{GOTPM1024 snapshots at redshift $z$\,=\,0.000 (left), $z$\,=\,1.012 (middle), and $z$\,=\,3.106 (right) with thickness of $10\Mpc$ in the $Z$-direction. The color bar indicates the environmental parameter ($\Phi_{\rm Env}$), which is the percentile rank of the enclosed total mass of halos with masses greater than $10^{9.8}\hMsun$ within a comoving radius of $5\Mpc$. The size of circles represents five times the virial radius. \label{fig:snap}}
\end{figure*}

\subsection{Halo Pair Sample}

The distance between two halos in a pair is the most important factor in identifying the interacting pair sample. 
Interactions happen when a halo is affected by the gravitational force from another halo (noncontact interactions) and when their virial regions of halos overlap each other (contact interactions).
In this paper, we take into account only the contact interaction case, whose distance criterion is expressed by
\begin{equation}
D_{\rm 12} < R_{\rm vir,1} + R_{\rm vir,2} , 
\end{equation}
where $D_{\rm 12}$ is the distance between the center positions of two constituent halos and ${R}_{\rm vir}$ is the virial radius of each halo. 
As mentioned in Section 2, the halo center is defined as the center of mass of the bound particles in the halo's innermost region to avoid misinterpretation by the tidally perturbed structures during the contact interaction.

This study focuses on the major interaction between two halos with similar masses (e.g., Milky Way and Andromeda).
We define the major interaction as the mass ratio ($M_{\rm neigh}/M_{\rm target}$) of halo pairs ranging from 1/3 to 3, where $M_{\rm target}$ is the mass of the target halo and $M_{\rm neigh}$ is the mass of its neighboring halo. 
The mass range of the target halos we deal with is $10^{10.8}$\,--\,$10^{13.0}\hMsun$.
The minimum mass ($10^{10.8}\hMsun$) of a target halo is equivalent to $\sim$400 member particles, and its neighboring halo with one-third of $10^{10.8}\hMsun$ has over 130 particles.

\subsection{The Environmental Parameter}

There are three main methods for defining the environment \citep{Mul12}: (1) $n$th-nearest-neighbor, (2) fixed aperture, and (3) fixed annulus.
We apply the fixed aperture method to the environment definition because this method is well known to describe the large-scale environment (such as fields, filaments, and clusters) better than other methods \citep{Mul12}.
The number count of halos is replaced by the total mass of halos in order to include the effect of massive halos.
To obtain the environmental parameter ($\Phi_{\rm Env}$), we first measure the total mass ($M_{\rm Env}$) of halos in our halo catalog ($M_{\rm halo}>10^{9.8}\hMsun$) within a sphere of a comoving radius of $R$\,=\,$5\Mpc$. 
Then, $\Phi_{\rm Env}$ is defined as a percentile rank order of $M_{\rm Env}$ given by
\begin{equation} \label{eq:rank}
\Phi_{\rm Env,i} \equiv {\rm \bf rank} (M_{\rm Env,i}),
\end{equation}
and
\begin{equation} \label{eq:cmass}
\ M_{{\rm Env},\,i} = \sum M_{{\rm halo},\,j}\ {\rm for}\ D_{ij} < 5\Mpc,    
\end{equation}
where $D_{ij}$ is the distance between the $i$th target halo and its $j$th neighboring halo. The range of $\Phi_{\rm Env}$ is from 0 to 100.

Figure \ref{fig:snap} shows three snapshots of the projected halo distributions with thickness of $10\Mpc$ in $Z$-direction at three different redshifts. 
The percentile rank method has an advantage of tracing given overdense regions from high redshifts to the present. 
The highly ranked environment at $z$\,=\,0.0 is largely highly ranked at $z$\,=\,3.1. 
This thus makes us reduce the selection effect when we analyze the redshift evolution of interaction fractions.
For the analysis of the environmental dependence of the redshift evolution, we classify the environment into low-density ($\Phi_{\rm Env}$\,$\leq$\,35), intermediate-density (35\,$<$\,$\Phi_{\rm Env}$\,$\leq$\,65), and high-density ($\Phi_{\rm Env}$\,$>$\,65) environments.
The figure shows that the environmental parameter, $\Phi_{\rm Env}$, properly represents the cosmic web type, and it is consistent with other environment definitions (e.g., \citealt{LHu15}; see Appendix \ref{sec:AB}).

\subsection{Definition of the Flyby and Merger}

The total energy of an interacting pair is required to determine whether or not the halos are bound to each other by mutual gravitational potential \citep{Ton12, Mor13}. 
In this analysis, the halo pair is simply modeled as an isolated two-body system. 
The total energy is a sum of kinetic and potential energies and calculated by
\begin{equation}\label{toten}
E_{12} = M_{1}M_{2}\left\{\frac{1}{2} \frac{|\vect{V_{1}} - \vect{V_{2}} + H(z)\vect{R_{12}}|^2}{M_{1}+M_{2}} - \frac{G}{|\vect{R_{12}}|}\right\},
\end{equation}
where $M_{1}$ and $M_{2}$ are the masses of the two constituent halos, $\vect{V_{1}}$ and $\vect{V_{2}}$ are their peculiar velocities with respect to the center of mass of the system, $H(z)$ is the Hubble parameter at a given redshift, and $\vect{R_{12}}$ is the displacement vector between the centers of the two halos.
The Hubble flow, $H(z)\vect{R_{12}}$, is usually in the form of the recession velocity when calculating the relative velocity between the halos.

If the total energy of a pair system is positive (negative), the pair would be unbound (bound).
However, the total energy of the system is reduced by the dynamical friction during the contact interactions, and even the sign of the total energy can turn from positive to negative. 
\citet{Sin12} showed that $\sim$20\% of flybys finally merged after their first passage. 
For a more robust classification (i.e., flybys or mergers) of the interaction, we employ the capture criterion of \citet{Gne03}. 
The capture criterion is defined as
\begin{equation}
\Delta E=2.4\,\mu\,\frac{M_{\rm neigh}}{M_{\rm target}}\,\frac{\sigma_{\rm target}^4}{V_{\rm rel}^2+\sigma_{\rm target}^2}\,, 
\label{eq:fric}
\end{equation}
where $\mu$ is a reduced mass of the pair system, $\sigma_{\rm target}$ is a velocity dispersion of the target halo, and $V_{\rm rel}$ is a magnitude of relative velocity between two constituent halos.
Although Equation \eqref{eq:fric} is analytically derived from the assumption of an isothermal density profile of a halo, the number fraction of pair systems with $0\leq E_{12}<\Delta E$ is consistent with the finding of \citet{Sin12}. 
Approximately 20\,\% of target halos with interacting neighbors with $E_{12}\geq0$ have a total energy smaller than the capture criterion (see Appendix C), and they merge with their neighbors despite the positive total energy.
We label an interacting halo pair as a flyby or a merger when the total energy is greater than the capture criterion ($E_{12}$\,$\geq$\,$\Delta E$) or smaller than the criterion ($E_{12}$\,$<$\,$\Delta E$), respectively.
The analysis focuses only on the current status of pairs, so it is not necessary to consider the merging time scale with respect to the Hubble time.

\subsection{Interaction Fraction}

In this study, three interaction fractions are considered in the halo pair catalog: the interaction fraction ($F_{\rm I}$), the flyby fraction ($F_{\rm F}$), and the merger fraction ($F_{\rm M}$). 
The interaction fraction is defined as
\begin{equation}
F_{\rm I} \equiv{} \frac{N_{\rm I}}{N_{\rm target}}\,, 
\end{equation}
where 
\begin{equation}
N_{\rm I} = \sum_i^{N_{\rm target}} N_{ {\rm I},i}\,,
\end{equation}
and the interaction number of the $i$th halo is $N_{{\rm I},i}=0$ or 1.
We define $N_{{\rm I},i}=1$ even in the case that the $i$th target halo has multiple interacting (flybying or merging) neighbors.
Consequently, $F_{\rm I}$ ranges between 0 and 1.
This thus enables us to avoid the number of interactions to be scaled with $\mathcal{O}(N^2)$.
Likewise, we apply the same rule to the flyby and merger cases:
\begin{equation}
N_{\rm F} = \sum_i^{N_{\rm target}} N_{{\rm F},i}\ \  {\rm and}\ \ N_{\rm M} = \sum_i^{N_{\rm target}} N_{{\rm M},i}.
\end{equation}
Note that a halo may have flyby and merger events at the same time.
In the halo pair sample of $10^{10.8}\hMsun$\,$<$\,$M_{\rm target}$\,$<$\,$10^{13.0}\hMsun$ at $z=0$ extracted from 11 simulations, 30,392 halos have interacting neighbors, among which 14,693 halos have flybying neighbors and 17,155 halos have merging neighbors. 
About 4.8\,\% of the halo pair sample have flyby and merger events simultaneously at $z=0$. 
We will discuss the multiple-interaction cases in detail in Section 4.3.

\section{Results}

\subsection{Mass versus Environment Planes of \texorpdfstring{$F_{\rm I}$, $F_{\rm F}$, $F_{\rm M}$, and $F_{\rm F}/F_{\rm M}$}{}}

\begin{figure*}[h]
\centering
\includegraphics[width=\textwidth]{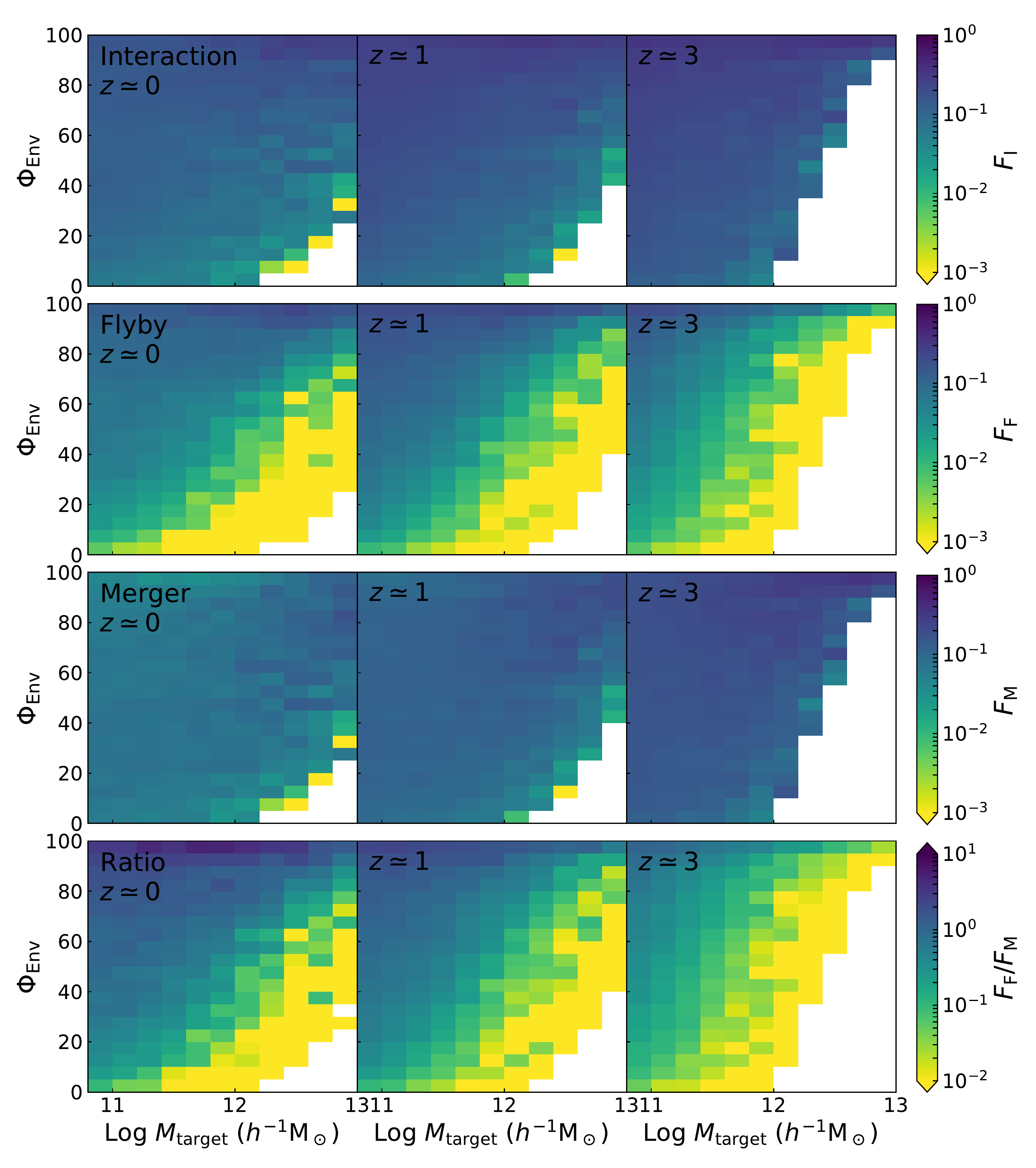}
\caption{Interaction fraction ($F_{\rm I}$; top row), flyby fraction ($F_{\rm F}$; second row), merger fraction ($F_{\rm M}$; third row), and ratio between $F_{\rm F}$ and $F_{\rm M}$ ($F_{\rm F}/F_{\rm M}$; bottom row) as functions of the halo mass and the environmental parameter ($\Phi_{\rm Env}$) at $z$\,$\simeq$\,0 (left columns), $z$\,$\simeq$\,1 (center), and $z$\,$\simeq$\,3 (right).
To enhance the statistical significance, the data from five successive snapshots are combined.
Four values are expressed by the color bars on the right, and bins with the number of target halos smaller than 50 are omitted and left as empty blocks. 
\label{fig:massenv}}
\end{figure*}

Figure \ref{fig:massenv} illustrates the interaction fraction ($F_{\rm I}$), flyby fraction ($F_{\rm F}$), merger fraction ($F_{\rm M}$), and ratio between $F_{\rm F}$ and $F_{\rm M}$ ($F_{\rm F}/F_{\rm M}$) as functions of the halo mass and environment at three different redshifts: $z$\,$\simeq$\,0 (left columns), $z$\,$\simeq$\,1 (center), and $z$\,$\simeq$\,3 (right). 
In the top row, the interaction fraction shows a marginal correlation with the halo mass and environment at all redshifts. 
The $F_{\rm I}$ values slightly decrease with increasing halo masses  and with decreasing $\Phi_{\rm Env}$. 
The trend is consistent qualitatively with that of \citet{LHu15}.
Whereas \citet{LHu15} used a mass ratio criterion of $M_{\rm neigh}/M_{\rm target}$\,$>$\,0.4 with no upper boundary, our criterion ($1/3<M_{\rm neigh}/M_{\rm target}<3$) sets an upper boundary, by which we do not count interactions of less massive target halos with more massive neighbors, which occur more frequently in the denser environment.
As a result, our mass and environmental dependence is weaker than that of \citet{LHu15}.

In the second row, the flyby fraction shows a clear dependence on the halo mass and environment at all redshifts of interest.
At a given environment (and redshift), $F_{\rm F}$ for the least massive halos is higher by up to a factor of 50 than for the most massive counterparts.
The mass dependence tends to diminish in denser environments.
While the anticorrelation between the flyby fraction and the halo mass is consistent with the interaction fraction and \citet{LHu15}, it is opposite to \citet{Sin12}, who showed a positive correlation.
We attribute the opposite behavior to \citet{Sin12} having a mass ratio criterion that is different from ours.
The flyby sample of \citet{Sin12} has no lower boundary in the mass ratio (i.e., $M_{\rm neigh}/M_{\rm target}$\,$<$\,1.0). 
With no lower boundary, the flyby fraction is boosted by numerous interactions of massive halos with smaller satellites and thus correlated with the halo mass.
At a given mass (and redshift), $F_{\rm F}$ increases by up to a factor of 400 with the increasing environmental parameter.
This environmental dependence is qualitatively consistent with observational surveys \citep[e.g.,][]{Lin10} and simulations \citep[e.g.,][]{Jia12}.

In the third row, the merger fraction shows a quite different trend from the flyby fraction in that the mass and environmental dependencies of $F_{\rm M}$ are very weak over all redshifts analyzed in this study.
The apparent mass independence of $F_{\rm M}$ is consistent with observations \citep{Xu12}. However, the environmental independence is at odds with the findings of \citet{Lin10} and \citet{Jia12}.
The origin of the mass and environmental dependence of $F_{\rm F}$ and the independence of $F_{\rm M}$ will be discussed in Section 5.1.

In the bottom row, we plot the $F_{\rm F} / F_{\rm M}$ ratio. 
The ratio depends on the halo mass and environment, analogously to the flyby fraction.
Obviously, this is because the flyby fraction only shows the clear dependence on these parameters, while the merger fraction does not.
The $F_{\rm F} / F_{\rm M}$ ratio increases with decreasing halo mass (by up to a factor of 100) and with the increasing $\Phi_{\rm Env}$ (by up to a factor of 300).
The maximum values of $F_{\rm F} / F_{\rm M}$ are 4.97, 2.49, and 1.12 at $z$\,$\simeq$\,0, 1, and 3, respectively.
In other words, the fractional contributions of flybys to the pair fraction are 0.83, 0.71, and 0.53 at $z$\,$\simeq$\,0, 1, and 3, respectively.
The flyby, as a source of tidal perturbations, starts to outnumber the merger ($F_{\rm F} / F_{\rm M}>1$) at high redshift ($z$\,$\sim$\,3) from the bins with the least halo mass in the highest density.
At $z\simeq 0$, the flyby outnumbers the merger in 58 bins among the total 220 bins ($\sim$26\,\%) on the mass--environment plane.

\begin{figure*}[htb!]
\centering
\includegraphics[width=0.93\textwidth]{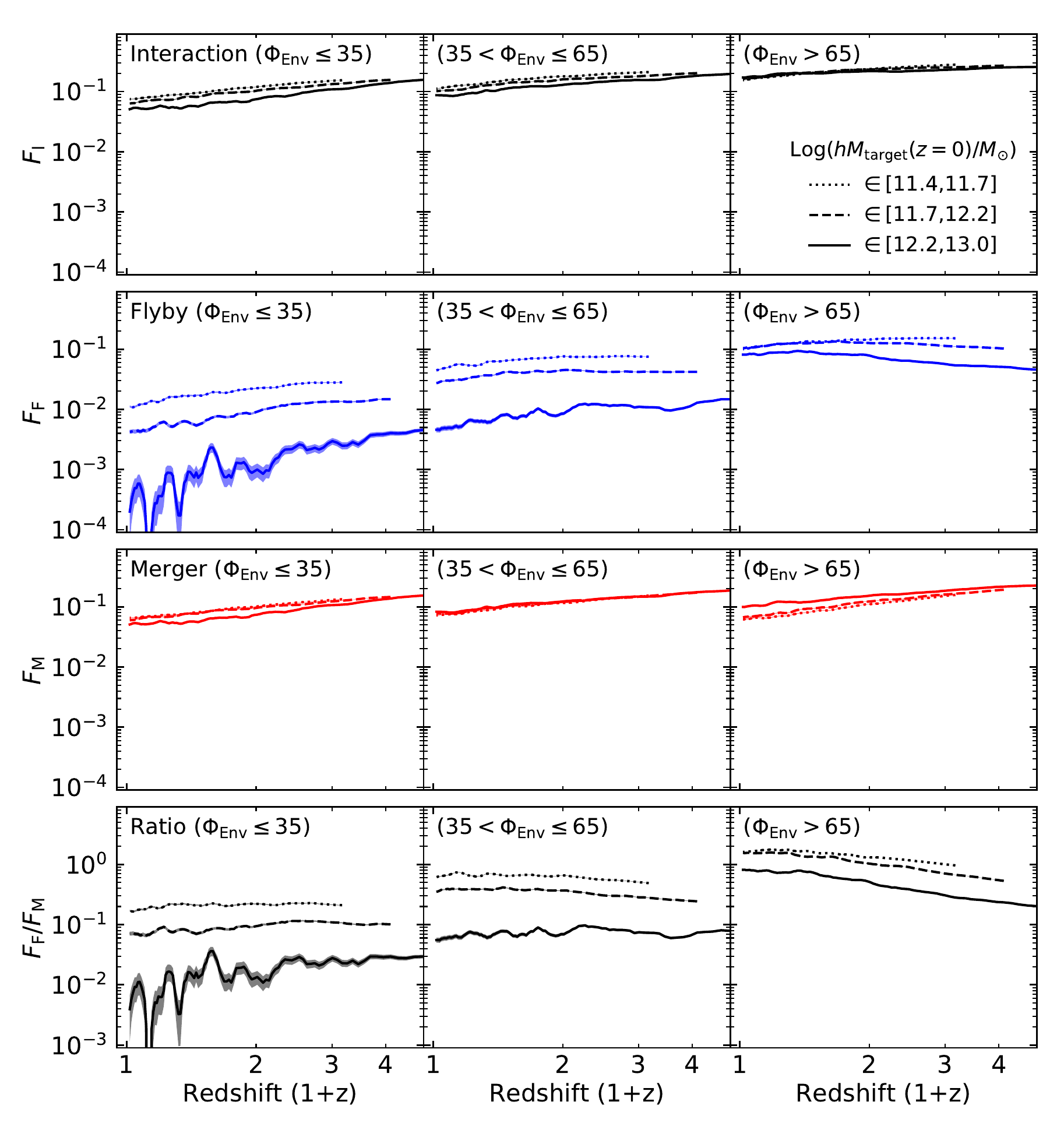}
\caption{Redshift evolution of the interaction fraction ($F_{\rm I}$; top row), flyby fraction ($F_{\rm F}$; second row), merger fraction ($F_{\rm M}$; third row), and ratio between $F_{\rm F}$ and $F_{\rm M}$ ($F_{\rm F}/F_{\rm M}$; bottom row) for the low-mass halos ($\log_{10}(hM_{\rm target}(z=0)/\Msun)$\,$\in$\,[11.4,\,11.7]; dotted lines), intermediate-mass halos ($\in$\,[11.7,\,12.2]; dashed lines), and high-mass halos ($\in$\,[12.2,\,13.0]; solid lines) in the low-density ($\Phi_{\rm Env}$\,$\leq$\,35; left), intermediate-density (35\,$<$\,$\Phi_{\rm Env}$\,$\leq$\,65; center), and high-density ($\Phi_{\rm Env}$\,$>$\,65; right) environments.
The Poisson error is shown by the shaded region. 
\label{fig:redevol}}
\end{figure*}

\subsection{Redshift Evolution of \texorpdfstring{$F_{\rm I}$, $F_{\rm F}$, $F_{\rm M}$, and $F_{\rm F}/F_{\rm M}$}{}}

Figure \ref{fig:redevol} presents the redshift evolution of the interaction fractions (top row), flyby fractions (second row), merger fractions (third row), and ratio of the flyby fraction to the merger fraction (bottom row).
Three halo mass subsamples are compared in the range of $10^{11.4}\hMsun$\,$<$\,$M_{\rm target}(z=0)$\,$<$\,$10^{13.0}\hMsun$ in three different (low-, intermediate-, and high-density) environments. 
To take into account the progenitor bias \citep[e.g.,][]{vanD01}, we set the halo mass range as a function of redshift by adopting the halo mass growth function from \citet{Cor15}:
\begin{equation} \label{zevol}
M_{\rm target}(z) = M_{\rm target}(z=0)\,(1+z)^{0.24}\,e^{-0.75z}.
\end{equation}
The progenitor halos of $M_{\rm target}(z=0)$ have a mass of $M_{\rm target}(z)$. 
By using this method, we can estimate the past interaction fraction of halos with $M_{\rm target}(z=0)$.
An additional mass cut at each redshift should be considered to have its corresponding halo mass equal to $10^{10.8}\hMsun$ at the starting redshift because the minimum halo mass is set to be $M_{\rm min} = 10^{10.8}\hMsun$. 
According to Equation \ref{zevol}, we divide the halo pair sample into three subsamples: high-mass halos ($\log_{10}(hM_{\rm target}(z=0)/\Msun)$\,$\in$\,[12.2,13.0]), intermediate-mass halos ($\in$\,[11.7,12.2]), and low-mass halos ($\in$\,[11.4,11.7]).
The lower boundary masses of the subsamples become greater than the minimum mass at $z=2.2$, $3.1$, and $5.0$, respectively.

In the top row of Figure \ref{fig:redevol}, the interaction fractions increase with redshift, regardless of the halo mass and environment.
On average, the $F_{\rm I}(z=0)$ values are 0.06, 0.10, and 0.16 and the $F_{\rm I}(z=2)$ values are 0.13, 0.18, and 0.25, for the low-, intermediate-, and high-density environments, respectively.
The slope seems slightly different depending on the environment.
The slopes for the low-, intermediate-, and high-mass halos are steeper by a factor of $\sim$1.2, $\sim$1.7, and $\sim$2.6 in the low-density environment than in the high-density environment.

In the second row, the flyby fraction evolves depending on the environment.
In the low-density environment, the flyby fraction increases with redshift.
The increasing trend gets stronger for more massive halos: $F_{\rm F}$ at $z$\,=\,2 for the high-mass (low-mass) halos is $\sim$15 ($\sim$2.5) times higher than that at $z$\,=\,0.
In the intermediate-density environment, the redshift evolution of $F_{\rm F}$ is qualitatively consistent with that in the low-density environment. 
Note that $F_{\rm F}$ evolves weakly with redshift at $z<1$, but remains constant at $z>1$.
In the high-density environment, $F_{\rm F}$ peaks at $z$\,$\sim$0.35 ($\sim$0.65), and the maximum value of $F_{\rm F}$ is 0.13 (0.09) for the high-mass halos (intermediate-mass halos). Then it slightly decreases with redshift.
In the third row, the merger fraction monotonically increases with redshift.
The value of $F_{\rm M}$ is roughly 0.06 at $z=0$ and 0.15 at $z=2$.

The evolutionary trends of $F_{\rm I}$ and $F_{\rm M}$ are well described by a form of $(1+z)^m$ \citep[e.g.,][]{Lin10, Xu12} at all redshifts.
The value of $m$ is roughly 1.0 in three interaction fractions in spite of a transition in the slope of $F_{\rm F}$.
This weak redshift evolution with small $m$-value was recently reported for the extension of the galaxy pair sample to high redshift \citep[$z$\,$>$\,1; e.g.,][]{Kee14,Mun17}, in contrast to the strong trend at low redshift \citep[$z$\,$<$\,1; e.g.,][]{Con09,Gen09}.

In the bottom row of Figure \ref{fig:redevol}, the redshift evolution of $F_{\rm F} / F_{\rm M}$ behaves in different ways depending on the environment.
In the low-density environment, $F_{\rm F} / F_{\rm M}$ increases with redshift, and the slope tends to be steeper for more massive halos (by up to 10), similarly to $F_{\rm F}$.
The increasing trend diminishes in the intermediate-density environment and even gets reversed in the high-density environment.
This indicates that, as we approach the present epoch, the relative effect of the flyby gets stronger (weaker) in the high-density (low-density) environments.
Interestingly enough, in the high-density environment, $F_{\rm F} / F_{\rm M}$ for the low-mass halos exceeds one for the majority of the Hubble time. 
The value converges to 1.6 at $z=0$, indicating that 62\,\% of pairs are flybys.
The $F_{\rm F} / F_{\rm M}$ ratio for the intermediate-mass halos has also been over one since $z$\,$\sim$1.3.
For the high-mass halos, $F_{\rm F} / F_{\rm M}$ touches unity at the present epoch. 
In the high-density environment, flybys are as frequent as or more frequent than mergers at $z\lesssim2$.
We emphasize that the flyby fraction can affect the amplitude and slope of the observed pair fraction, varying with redshift.
\citet{Sny17} found that the number fraction of real mergers among galaxy pairs decreases from 0.8 at $z$\,=\,3 to 0.5 at $z$\,=\,1.
This is in line with our result that $F_{\rm F} / F_{\rm M}$ in the high-density environment becomes greater approaching $z=0$, and it suggests that the flyby fraction contributes more to the pair fraction at lower redshift.
We will discuss how significantly the flyby fraction contaminates the observed pair fraction in Section 5.3.

\begin{figure*}[htb!]
\centering
\includegraphics[width=0.93\textwidth]{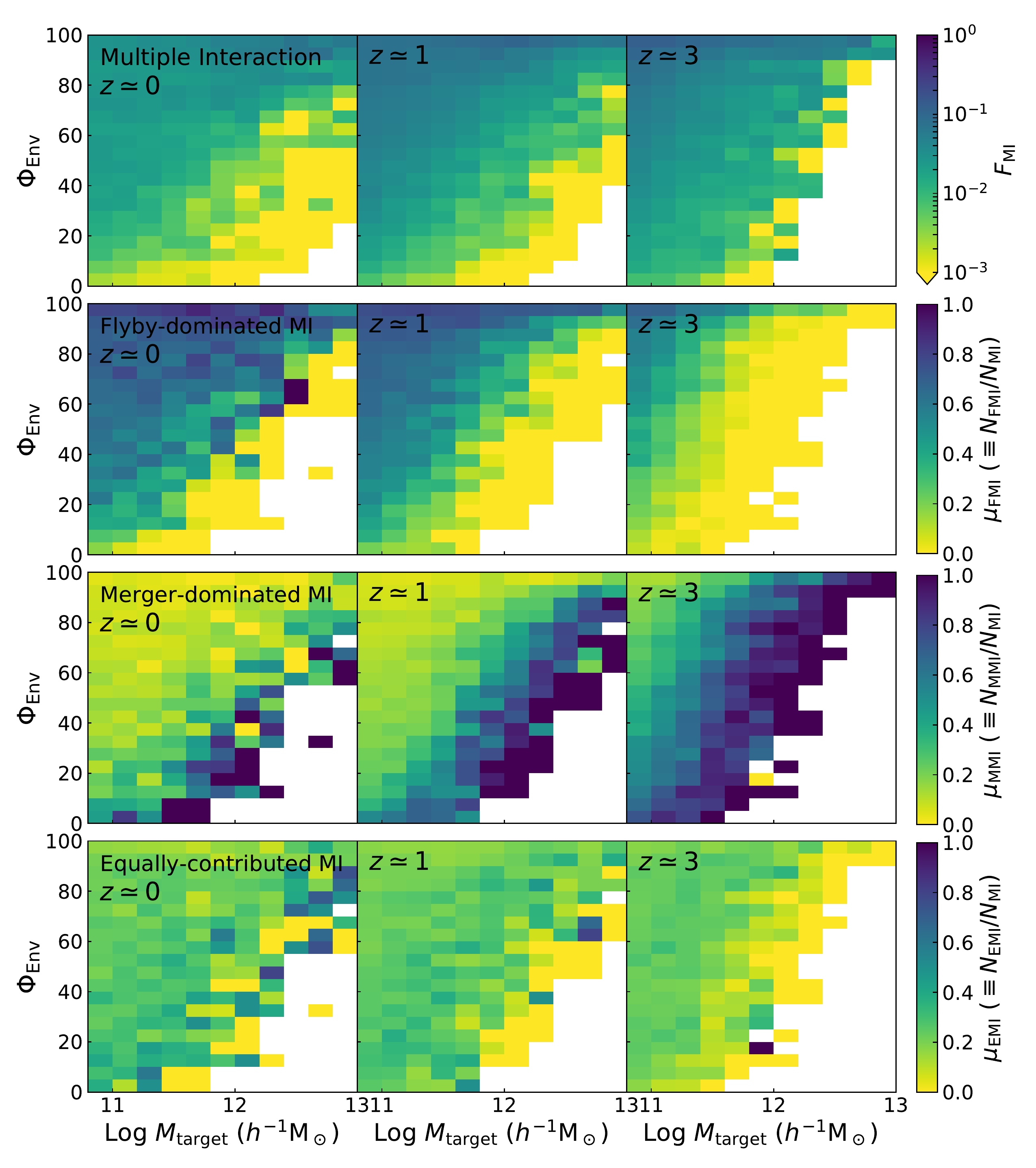}
\caption{Same as Figure \ref{fig:massenv}, but for the multiple interaction (MI) fraction ($F_{\rm MI}$; top row) and the relative contribution of flyby-dominated MIs ($\mu_{\rm FMI}$; second row), of merger-dominated MIs ($\mu_{\rm MMI}$; third row), and of equally contributed MIs ($\mu_{\rm EMI}$; bottom row) to the MI fraction. 
In all panels, bins with the number of target halos smaller than 50 are omitted and left as white, empty blocks. 
In the lower three panels, bins with $F_{\rm MI}=0.0$ (no multiple interaction) are also marked as empty blocks.
\label{fig:mimenv}}
\end{figure*}

\subsection{Multiple Interaction Fraction}

A target halo can have two or more interacting neighbors at the same time.
In our simulations, the number of target halos with two or more comparable-mass (i.e., 1/3\,--\,3) companions is 4557 ($\sim$\,15\,\% of $N_{\rm I}$, the total number of target halos with interacting neighbors) at $z$\,=\,0.
About 5\,\% of $N_{\rm I}$ possess both flybying and merging neighbors (see Section 3.4), and the remaining $\sim$ 10\,\% have multiple flybying neighbors only or multiple merging neighbors only. 
In this subsection, we measure the multiple interaction (MI) fraction, defined by
\begin{equation}
F_{\rm MI} \equiv{} \frac{N_{\rm MI}}{N_{\rm target}}\,, 
\end{equation}
where $N_{\rm MI}$ is the number of target halos with multiple interacting neighbors.
The multiple interactions are classified into three subsamples: (a) the flyby-dominated multiple interaction (FMI), in which the cumulative number of flybying neighbors is larger than that of merging neighbors ($n_{{\rm F},i}>n_{{\rm M},i}$); (b) the merger-dominated multiple interaction (MMI; $n_{{\rm F},i}<n_{{\rm M},i}$), which is opposite to the FMI; and (c) the equally-contributed multiple interaction (EMI), in which the numbers of flybying neighbors and merging neighbors are equivalent ($n_{{\rm F},i}=n_{{\rm M},i}$). Note that $n_{{\rm F},i}\ne N_{{\rm F},i}$; $n_{{\rm F},i}$ is a cumulative number, but $N_{{\rm F},i}$ is a noncumulative number (0 or 1). 

Figure \ref{fig:mimenv} presents the multiple interaction (MI) fraction ($F_{\rm MI}$; top row) and the contribution of the three types of multiple interactions to the MI fraction (second, third and bottom rows) on the halo mass--environment planes at three different redshifts.
In the top row, the MI fraction strongly depends on both the halo mass and environmental parameter at all redshifts, and the trend is similar to that of the flyby fraction (see second row of Figure \ref{fig:massenv}).
As the halo mass increases and $\Phi_{\rm Env}$ decreases, $F_{\rm MI}$ becomes lower by up to 40 and 100, respectively.
On average, $F_{\rm MI}$ increases with redshift.
The maximum values of $F_{\rm MI}$ are 0.078, 0.109, and 0.128 at $z\simeq0$, 1, and 3, respectively.
The MI fraction derived from our simulations tends to be higher than the observational result of \citet{Dar11}. 
They identified the multimerger system using the Galaxy Zoo catalog and found that the ratio of triple systems to binary systems is smaller than 2\% (multiple/binary\,$\leq$\,2.5\,\%). 
In our simulations, the ratio of $F_{\rm MI}$ to $F_{\rm I}$ at $z\simeq0$ is on average 0.15 (multiple/binary\,$\simeq$\,18\,\%), which is over five times higher than the observational upper limit.
This is because of our larger distance criterion (approximately 10 times higher than that of \citealt{Dar11}).
If we set the distance criterion to $30\kpc$, the ratio $F_{\rm MI}/F_{\rm I}$ is reduced to 0.045 (multiple/binary\,$\simeq$\,4.7\,\%) and is still two times higher than the result of \citet{Dar11}.
This is attributed to the underestimation of binary systems (binary/single\,$\sim$\,0.5\,\% in the case of $30\kpc$) by the low force resolution of our simulations ($\epsilon=12.5\kpc$).

\begin{figure*}[htb!]
\centering
\includegraphics[width=0.9\textwidth]{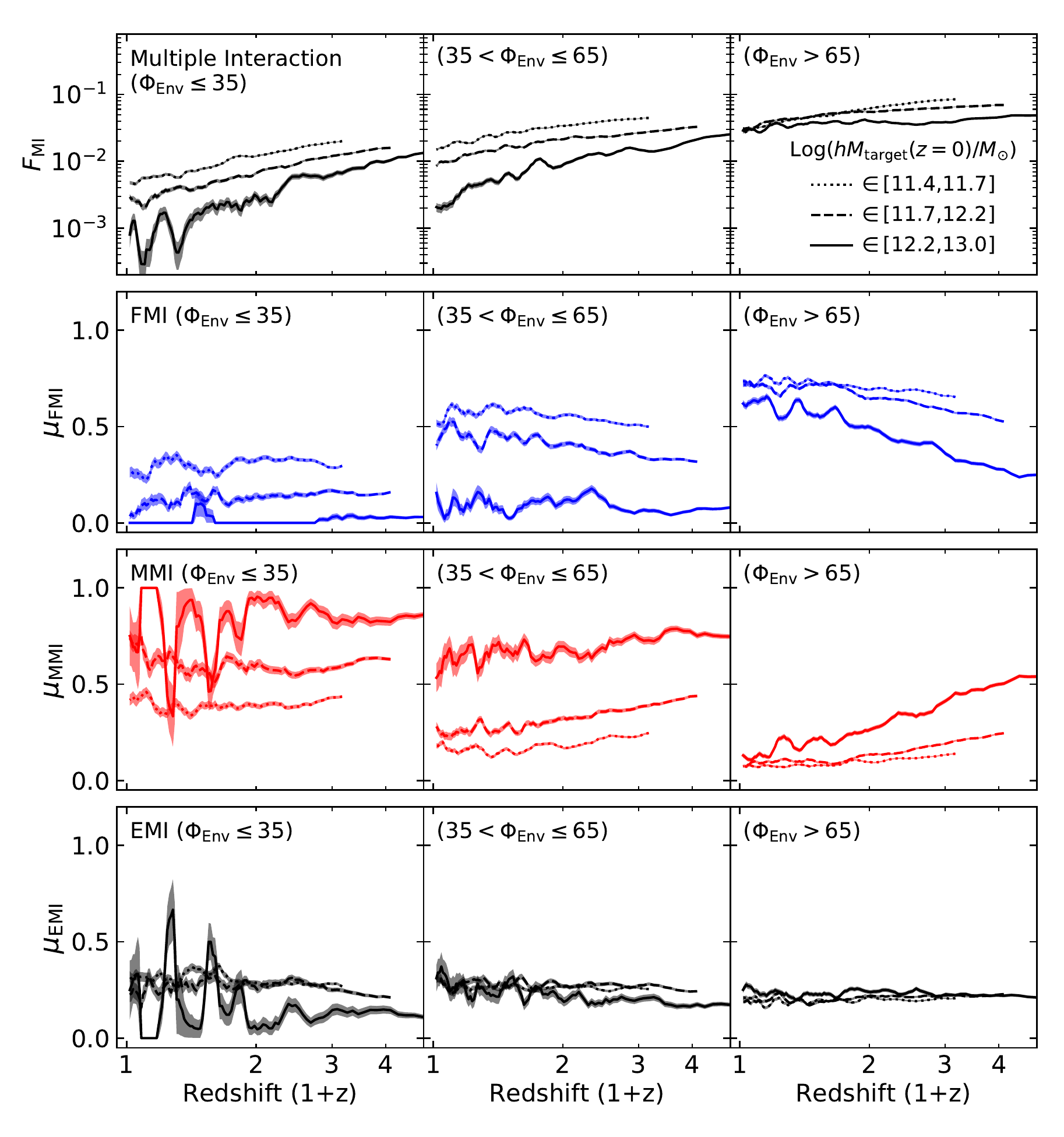}
\caption{Same as Figure \ref{fig:redevol}, but for the multiple interaction (MI) fraction ($F_{\rm MI}$; top row) and the relative contribution of flyby-dominated MI ($\mu_{\rm FMI}$; second row), of merger-dominated MI ($\mu_{\rm MMI}$; third row), and of equally contributed MI ($\mu_{\rm EMI}$; bottom row) to the MI fraction. 
\label{fig:mired}}
\end{figure*}

The three types of multiple interactions contribute differently to the MI fraction.
In the lower three rows of Figure \ref{fig:mimenv}, we plot the number fractions of multiple interaction types with respect to the number of MI target halos: (1) the number fraction of FMI target halos ($\mu_{\rm FMI}\equiv N_{\rm FMI}/N_{\rm MI}$; second row), (2) the number fraction of MMI target halos ($\mu_{\rm MMI}\equiv N_{\rm MMI}/N_{\rm MI}$; third row), and (3) the number fraction of EMI target halos ($\mu_{\rm EMI}\equiv N_{\rm EMI}/N_{\rm MI}$; bottom row).
The contribution of FMIs is both mass dependent and environment dependent (second row), similarly to the flyby fraction and MI fraction.
For the low-mass halos in the high-density environment, the majority of the MIs are FMIs.
Moreover, $\mu_{\rm FMI}$ tends to decrease with redshift.
The maximum contributions of FMIs are 1.0, 0.79, and 0.69 at $z\simeq0$, 1, and 3, resepctively.
This is in line with the trend of $F_{\rm F}/F_{\rm M}$ in the high-density environment (see bottom row of Figure \ref{fig:redevol}), where flybys happen the most frequently.
The contribution of MMI ($\mu_{\rm MMI}$) shows an opposite trend to that of FMI (third row).
As the halo mass increases and $\Phi_{\rm Env}$ decreases, $\mu_{\rm MMI}$ becomes higher, and these conditions cause the merger fraction to highly exceed the flyby fraction (see bottom row of Figure \ref{fig:redevol}).
In the bottom row, the EMI contribution ($\mu_{\rm EMI}$) has weaker mass and environmental dependence, compared to $\mu_{\rm FMI}$ and $\mu_{\rm MMI}$.
The mean value of $\mu_{\rm EMI}$ is about 0.23.
Only for halos with $M_{\rm target}>10^{12}\hMsun$ does the mass dependence emerge; $\mu_{\rm EMI}$ tends to decrease with increasing halo masses.

To further investigate differences in $F_{\rm MI}$, $\mu_{\rm FMI}$, $\mu_{\rm MMI}$, and $\mu_{\rm EMI}$, we plot the redshift evolution of four variables for nine subsamples in Figure \ref{fig:mired}.
The top row shows that the MI fraction rises with redshift.
For instance, for the intermediate-mass halos, the values of $F_{\rm MI}$ at $z=0$ ($z=2$) are 0.003, 0.009, and 0.029 (0.012, 0.028, and 0.062) in the low-, intermediate-, and high-density environments, respectively.
In the low- and intermediate-density environments, the evolutionary trend is steeper for the higher mass halos, while it is reversed in the high-density environment. 
Compared to the interaction fraction, the MI fraction on average evolves two times more rapidly with redshift.

\startlongtable
\begin{deluxetable*}{lcccccc}
\tablecaption{Summary of the Number of Target Halos with Interacting Neighbors at $z=0$. \label{tab:nneigh}}
\tablewidth{0pt}
\tablehead{
\multirow{2}{*}{Sample} &
\multirow{2}{*}{$N_{\rm target}$} & \multirow{2}{*}{$N_{\rm F}$} &
\multirow{2}{*}{$N_{\rm M}$} & \multicolumn{3}{c}{$N_{\rm MI}$}\\
\cline{5-7}
&  &  &  & \colhead{all} & \colhead{$n_{\rm F}>n_{\rm M}$} & \colhead{$n_{\rm M}>n_{\rm F}$}} 
\decimals
\startdata
All & 263,256 & 14,693 & 17,155 & 4557 & 2841 & 588 \\
\cline{1-7}
Low-mass & 35,304 & 1865 & 2316 & 561 & 352 & 81 \\
Intermediate-mass & 26,444 & 1287 & 1800 & 426 & 282 & 65 \\
High-mass & 11,682 & 455 & 1003 & 176 & 105 & 30 \\
\cline{1-7}
Low-density & 92,888 & 1893 & 6027 & 744 & 313 & 200 \\
Intermediate-density & 79,184 & 4012 & 5668 & 1371 & 808 & 189 \\
High-density & 91,184 & 8788 & 5460 & 2442 & 1720 & 199 \\
\enddata
\tablecomments{``All'' sample includes all target halos with $\log_{10}(hM_{\rm target}/\Msun)$\,$\in$\,[10.8,\,13.0] and $0<\Phi_{\rm Env}\leq100$. The definitions of our mass and environment subsamples are specified in Sections 4.2 and 3.2, respectively.}
\end{deluxetable*}

In the lower three rows, the contributions of the three types of multiple interactions evolve differently with redshift depending on the environment.
In the low-density environment, three contributions are almost independent of redshift. 
In most ranges of redshift, $\mu_{\rm MMI}$ keeps higher values than $\mu_{\rm FMI}$ and $\mu_{\rm EMI}$.
The values of $\mu_{\rm MMI}$ are approximately 0.39, 0.60, and 0.85 for the low-, intermediate-, high-mass halos.
In the intermediate-density environment, $\mu_{\rm MMI}$ slightly increases with redshift while $\mu_{\rm FMI}$ and $\mu_{\rm EMI}$ marginally decrease.
The MI type that contributes the most depends on the halo mass: $\mu_{\rm FMI}$ for the low-mass halos and $\mu_{\rm MMI}$ for the high-mass halos. 
For the intermediate-mass halos, $\mu_{\rm FMI}$ is the highest at $z<1.5$ but $\mu_{\rm MMI}$ at $z>1.5$.
In the high-density environment, three contributions show different evolutionary trends. 
At all redshifts, $\mu_{\rm FMI}$ decreases, $\mu_{\rm MMI}$ increases, and $\mu_{\rm EMI}$ remains constant.
The FMI is mostly the MI type with the highest contribution, except for the high-mass halos at $z>2$ where $\mu_{\rm MMI}$ is higher than $\mu_{\rm FMI}$.
At $z=0$, $\mu_{\rm FMI}$ is roughly 0.7.

Table \ref{tab:nneigh} gives the number of target halos with flybying and merging neighbors. 
The number of target halos with flybying neighbors is in general smaller than that with merging neighbors except for the high-density environment. 
However, in the case of the multiple interaction, the $n_{\rm F}>n_{\rm M}$ cases always outnumber the $n_{\rm M}>n_{\rm F}$ cases.
Even in the high-density environment, $\sim$\,20\,\% of target halos with flybying neighbors have multiple flybying neighbors, and thus the frequency of the flyby is about two times greater than that of the merger.

\section{Discussion} 

\subsection{Why Does the Flyby Fraction Only Depend on Mass and Environment?}

In this subsection, we delve into the physical causes of the mass and environmental dependence of the flyby fraction and the independence of the merger fraction. 
We consider the following three factors: (1) the possibility of running into a halo with a comparable mass (i.e., an encounter with a mass ratio of 1:3\,--\,3:1), which is higher for lower mass halos; (2) the relative velocity of two halos in a pair, which depends on their large-scale environment; and (3) the depth of the gravitational potential well of a halo, which depends on the mass of the halo.

First, the flyby fraction has an anticorrelation with the halo mass. 
According to the halo mass function, the number density of less massive halos is higher than that of more massive halos \citep[e.g.,][]{Pre74}.
Less massive halos have higher possibility of an encounter with halos of comparable mass and thus have the higher flyby fraction.
For the same reason, the multiple interaction fraction is also higher for the less massive halos.
The merger fraction, on the other hand, is rather independent of the halo mass. 
This is because there are two competing factors regulating the merger fractions: less massive halos tend to encounter more frequently comparable-mass halos but have difficulty at the same time in holding nearby halos, due to their shallower potential well.
Many studies have shown the diverse results of the mass dependence of the merger fraction \citep{Gen09,Xu12,Cas14,Kee14,Rod15}.
The diversity can be accounted for by the difference in the mass ratios at the infall moment and right before the merger \citep[e.g.,][]{Rod15} and the diverse definitions of the major merger \citep[][and references therein]{Cas14}.

\begin{figure*}[ht!]
\includegraphics[width=0.9\textwidth]{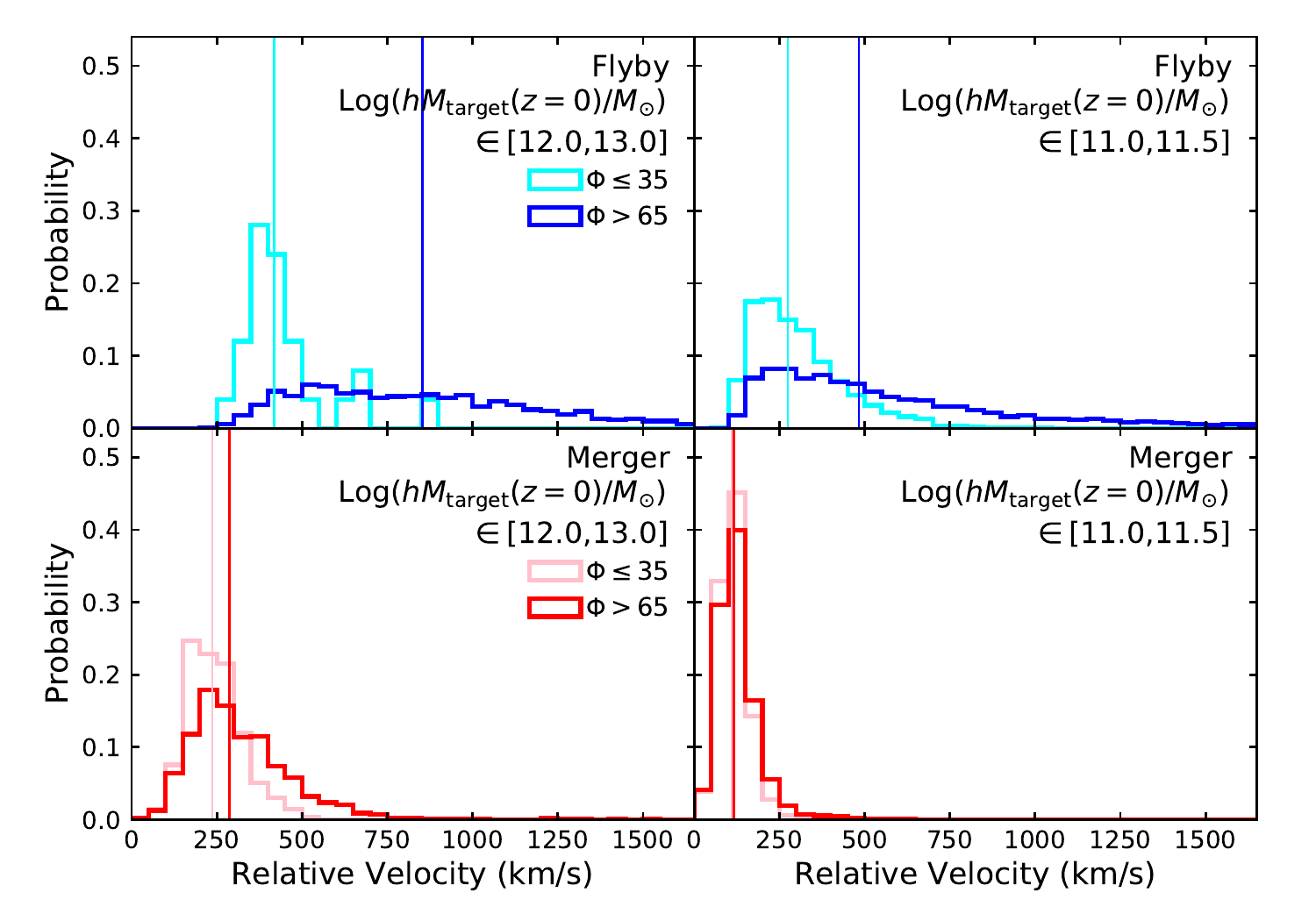}
\centering
\caption{Relative velocity distributions of flybys (upper) and mergers (lower) at $z$\,$\sim$\,0 for the higher mass halos ($\log_{10}(hM_{\rm target}(z=0)/\Msun)$\,$\in$\,[12.0,\,13.0]; left) and lower mass halos ($\in$\,[11.0,\,11.5]; right).
The cyan and pink histograms are for the halo pairs in the low-density environment ($\Phi_{\rm Env}$\,$\leq$\,35) and the blue and red histograms are for the pairs in the high-density ($\Phi_{\rm Env}$\,$>$\,65) environment.
The pairs are extracted from five successive snapshots near $z$\,=\,0. 
The vertical lines indicate the median values of relative velocities in the low- and high-density environments. \label{fig:frelvel}}
\end{figure*}

Next, the flyby fraction has a positive correlation with the environmental parameter. 
We attribute this to the high velocity dispersion in denser environments \citep[e.g.,][]{Evr08}.
Figure \ref{fig:frelvel} shows the relative velocity distributions of flybying neighbors (upper) and merging neighbors (lower).
For the flybying neighbors in the upper panels, the median relative velocity for the higher mass (lower mass) halos is $\sim$420 ($\sim$270)$\kms$ and $\sim$850 ($\sim$480)$\kms$ in the low- and high-density environments, respectively.
Intuitively, in the high-density environment, the relative velocity exceeds more easily the escape velocity, and thus the number of flybying neighbors is enhanced.
In addition to the high relative velocity, the flyby fraction is higher because the possibility of encountering another halo is higher in denser environments.
As the high relative velocity and the high possibility arise at the same time, the flyby fraction becomes more strongly dependent on the large-scale environment.
For the merging neighbors in the lower panels, the relative velocity distributions do not depend on the environment but on the halo mass.
The median velocity for the higher-mass (lower-mass) halos is $\sim$\,$260$\,(110)$\kms$ regardless of the environment.
The environmental independence of the median relative velocity explains why the merger fraction shows little or no environmental dependence.
Despite the higher encounter possibility in denser environments, most relative velocities of interacting neighbors exceed the escape velocity of their target halos, lowering the merger fraction.
This is in line with the notion that it is hard for a merger to happen in the cluster environment \citep{Ost80,Wet08}.

\begin{figure*}[ht!]
\includegraphics[width=0.9\textwidth]{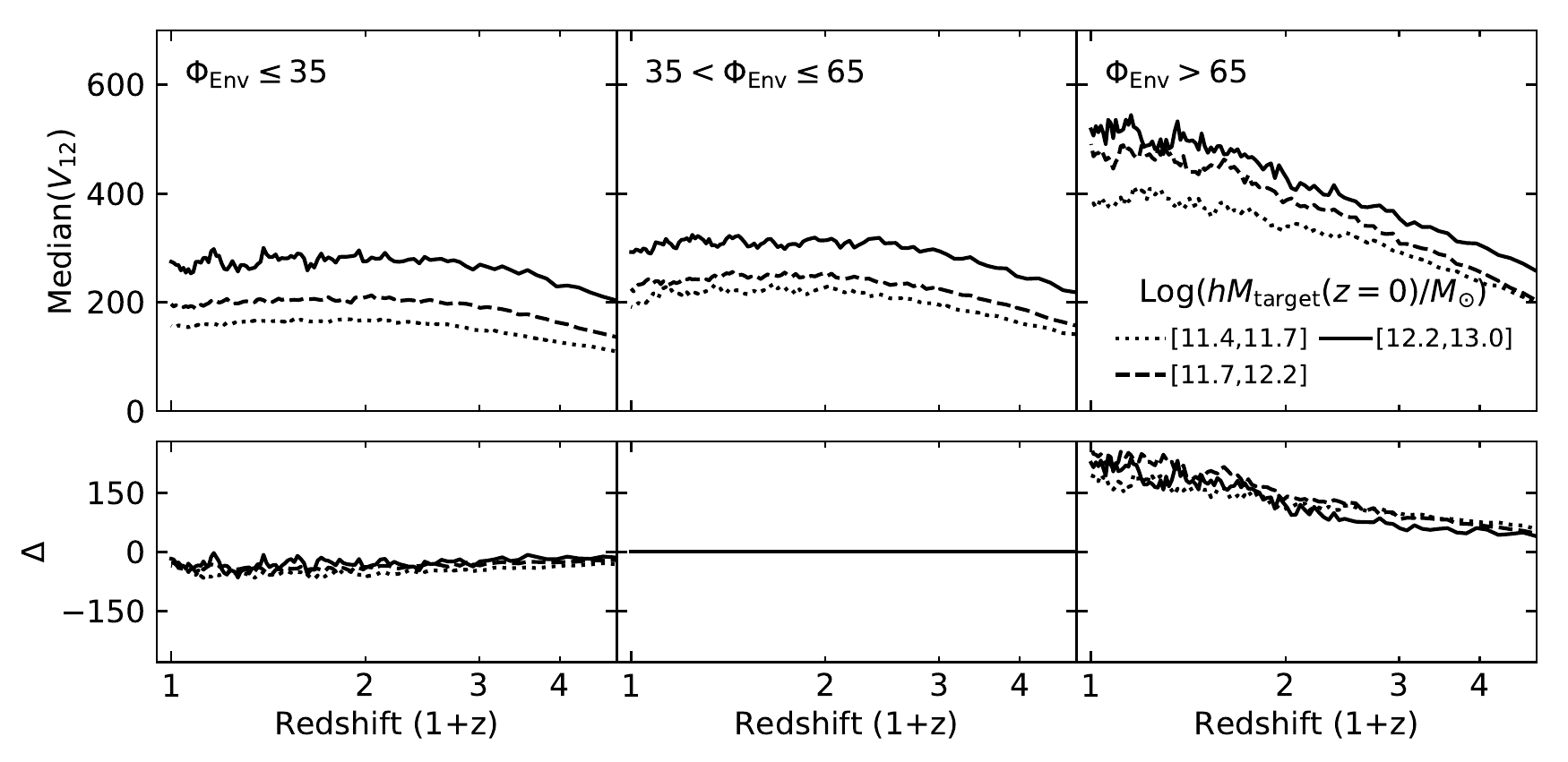}
\centering
\caption{Upper panels: The median value of relative velocities of interacting companions (both flybys and mergers) as a function of redshift. Line styles indicate three different halo mass samples: the low-mass halo (dotted), the intermediate-mass halo (dashed), and the high-mass halo (solid). Lower panels: The difference ($\Delta$) in the median value between three different environments and the intermediate-density environment for the same mass bins. Therefore, all values are zero in the bottom center panel. 
\label{fig:medrv}}
\end{figure*}

\subsection{Interpretation of Redshift Evolution of \texorpdfstring{$F_{\rm F}$, $F_{\rm M}$, and $F_{\rm F}/F_{\rm M}$}{}}

In this subsection, we interpret the redshift evolution of the flyby fraction, the merger fraction, and their ratio in the context of the formation history of the cosmic large-scale structure.
Figure \ref{fig:medrv} shows the median relative velocities of all interacting pairs (both flybys and mergers) as a function of redshift.
In the low- and intermediate-density environments, the median relative velocity slightly increases from $z$\,=\,4 to $\sim$\,1.
The enhanced relative velocity hinders the flyby fraction from decreasing, similarly to the merger fraction.
But after $z\simeq1$ the median relative velocity is roughly constant, and at the same time the flyby fraction starts to decrease (see second row of Figure \ref{fig:redevol}).
In contrast, in the high-density environment, the median relative velocity dramatically increases toward $z$\,=\,0.
We attribute the enhancement of the flyby fraction from $z$\,=\,4 to $\sim$\,0.3 in the high-density environment to the dramatic rise in the median relative velocity.
This effect also leads to both the enhanced contribution of flyby-dominated multiple interactions and the reduced contribution of merger-dominated multiple interactions with respect to the multiple interaction fraction.
Furthermore, the gap of the median values between the low-density and the high-density environments gets wider toward $z$\,=\,0, indicating that the environmental dependence of the ratio $F_{\rm F} / F_{\rm M}$ becomes stronger when approaching $z$\,=\,0.
The boosted environmental dependence is in line with the evolution of the cosmic web \citep[from voids to nodes;][]{Bon96,Cau14}.
According to \citet{Cau14}, as the total volume of voids expands over cosmic time, the relative velocity between two halos gets relatively lower, and accordingly $F_{\rm F} / F_{\rm M}$ gets reduced. By contrast, as the total volume of nodes contracts over cosmic time, $F_{\rm F} / F_{\rm M}$ gets enhanced.

\begin{figure*}[ht!]
\includegraphics[width=0.9\textwidth]{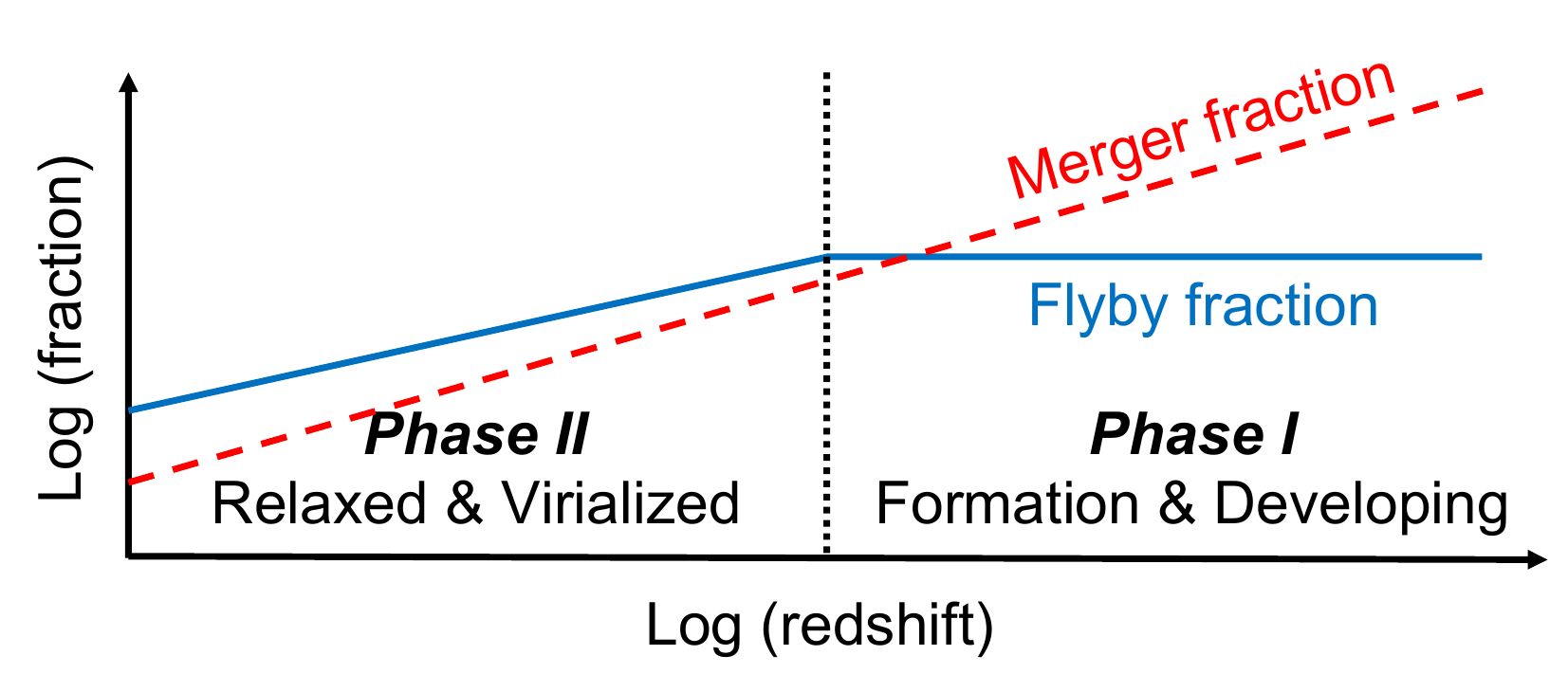}
\centering
\caption{Simple schematic diagram of the redshift evolution of the flyby fraction (blue solid line) and the merger fraction (red dashed line). 
The cosmic structural history of the halo system is divided into two phases and linked to the evolutionary trends of the flyby and merger fractions. 
The vertical dotted line indicates the turnover redshift of the flyby fraction.
\label{fig:fbscenario}}
\end{figure*}

Figure \ref{fig:fbscenario} is a simple schematic diagram illustrating our scenario, which links the redshift evolution of two fractions to the hierarchical structure formation.
We divide the formation history into two phases based on \citet{Kne04,Kne06}, who found a correlation between the dynamical age of simulated clusters and the frequency of flybys: the younger, the higher.
First, in Phase I, protoclusters and protogroups formed mainly via major mergers in the early universe. 
At this epoch, because the relative velocity between halos was still low, the merger fraction was higher than the flyby fraction.
During this phase, as the large-scale structure was gradually growing, infalls of halos and groups took place more frequently with the increasing relative velocity. 
Hence, the flyby fraction remains constant or slightly increases. 
By contrast, the merger fraction decreased with the decreasing redshift because the hierarchical merging leads to decline in the halo accretion rate over cosmic time \citep[e.g.,][]{Mcb09}. For the same reason, the multiple interaction fraction gets lower toward low redshift.

Next, in Phase II, groups and clusters became dynamically virialized and relaxed, and the number of comparable nearby halos decreased.
From the beginning of Phase II, the flyby fraction started to decline. 
The turnover of the redshift evolution for the flyby fraction is logically consistent with \citet{Kne04,Kne06}.
In addition, the diversity of the turnover redshift of the flyby fraction is attributed to the fact that the assembly and virialization of halo systems are mass dependent \citep[e.g.,][]{Pow12} and environment dependent \citep[e.g.,][]{She04}. 
During this phase, the continuous mergers and the resultant reduction of the number of comparable neighbors decrease monotonically the merger fraction. 
This is more so in the accelerating universe where halo systems become more isolated with time \citep{Bus03,Beh12}.

\subsection{Contamination of the Observed Pair Fraction by the Flyby Fraction}

Given that the observed pair (flyby + merger) fraction is used to derive the merger rate, it is important to examine to what extent the pair fraction is contaminated by flybys that are observationally indistinguishable from merging pairs.
Back in Figure \ref{fig:frelvel}, even up to $\sim$\,90\,\% of the flybys in the low-density region have relative velocities smaller than $500\kms$, suggesting that it is hard to determine, by using only a certain velocity criterion, whether a pair is a flyby or a merger.

Among the three types of dependence analyzed in this paper, the mass and environmental dependence of the flyby fraction should affect the behavior of the observed pair fraction.
Survey observations revealed that the pair fraction does not depend on the stellar mass \citep[e.g.,][]{Xu12} but strongly depends on the environment \citep[e.g.,][]{Lin10}.
The absence of the mass dependence seems due to the strict merger definition ($D_{\rm 12,projected}$\,$\lesssim$\,$30\kpc$ and $V_{\rm 12,radial}$\,$\lesssim$\,$200\kms$) in an attempt to reduce the effect of spurious mergers, and thus the mass dependence of the flyby fraction barely contributes to the pair fraction.
The interaction fraction also shows little mass dependence because the merger fraction is dominated by the flyby fraction ($F_{\rm F} / F_{\rm M} \lesssim 0.1$) in the low-density environment, where the mass dependence of $F_{\rm F}$ is the most evident.
On the other hand, the environmental dependence of the pair fraction has been studied with a rather looser definition for mergers ($D_{\rm 12,projected}$\,$\lesssim$\,$200\kpc$ and $V_{\rm 12,radial}$\,$\lesssim$\,$1000\kms$) in order to search for paired companions in cluster environments.
Given that in dense environments, $F_{\rm F} / F_{\rm M} \gtrsim1$, the looser merger definition allows the pair fraction to be more contaminated by flybys, resulting in the observed dependence of the pair fraction on the environment.

Flybys also contaminate the redshift evolution of the observational pair fraction.
Depending on the definition of galaxy pairs, the redshift evolution is observed to be strong \citep[e.g.,][]{Con09} or weak \citep[e.g.,][]{Kee14} at $z$\,$\lesssim$\,1 \citep[see also][]{Wil11, New12, Man16, Mun17}.
The diverse results are ascribed to the contribution of flybys that are observationally indistinguishable from pairs, depending on the merger definition.
The pair fraction is expected to be more flattened if the flyby contaminates the pair fraction more significantly.
Hence, in order to derive the {\it real} merger rate from the pair fraction, the ratio of flybys to pairs (from 0.004 to 0.62; see bottom panels of Figure \ref{fig:redevol}) should be considered.

Assessing the contribution of interactions to galactic evolution is further complicated by the following two factors.
Some flybying companions will be turned into merging companions by dynamical friction. 
We have minimized the effect of dynamical friction by applying the capture criterion to the classification of flybys.
Despite this consideration, the real number of flybys that a galaxy experiences during its lifetime, requires the real flyby rate \citep[e.g.,][]{Sin12} and the flyby time scale (between the beginning and end of a contact interaction).
In addition to the dynamical friction, some halos experience multiple interactions \citep{Dar11}. 
Although we found that the multiple interaction is mainly governed by the flyby-dominated MI, it is uncertain whether the multiple interaction raises the flyby effect \citep[e.g., the restricted three-body problem;][]{Sal07} or lowers it \citep[e.g., more mergers due to dynamical friction;][]{Lac93,Jia08}.
According to our results, multiple interactions in denser environments are expected to enhance the flyby effect, due to the high relative velocity and numerous concurrent flybys ($\sim$\,20\,\% of the flyby fraction).

\section{Summary and Conclusion}

We have presented the flyby fraction, the merger fraction, and their ratio as functions of the halo mass, large-scale environment, and redshift. 
To examine the three dependencies, a series of cosmological $N$-body simulations with a mass resolution of $1.5 \times 10^{8} \hMsun$, is performed under the $\Lambda$CDM cosmology (WMAP 9-year).
We have focused on the interacting systems with halo masses $10^{10.8}$--$10^{13.0}\hMsun$, with mass ratios of 1:3--3:1, and with a separation smaller than the sum of virial radii of the two halos. 
Our main results are summarized as follows.

\begin{enumerate}[(1)]

\item 
The flyby fraction increases by a factor of 50 with decreasing halo mass and by a factor of 400 with increasing large-scale density, while the merger fraction does not show any significant correlation. 
Various surveys found that the observed pair fractions depend on the environment, similarly to the flyby fraction, but do not depend on the mass, similarly to the merger fraction. 
This may suggest that the flyby contaminates significantly the observed pair fraction, and the sample-selection criteria of surveys on environmental dependence allow a higher level of contamination by flybys compared to studies on mass dependence.

\item 
The flyby fraction increases from $z$\,=\,0 to $\sim$\,1, and then it remains constant or marginally decreases. 
The merger fraction shows a monotonic increase with redshift. 
Combined with the hierarchical galaxy formation theory, we propose a scenario for the redshift evolution of the flyby and merger fractions: (a) The flyby fraction remains constant during the formation and developing phase of the large-scale structure, and then falls off after the relaxation and virialization of the structure; (b) As the halo systems hierarchically merge, the reduction of the number of neighbors leads to a decline of the merger fraction.

\item 
The multiple interaction fraction depends highly on the halo mass, environment, and redshift, similarly to the flyby fraction. 
Classifying the multiple interactions into three MI types, we find that the flyby-dominated MIs significantly contribute to the MI fraction, and the trend is opposite for the merger-dominated MIs.
The numerous concurrent flybying neighbors reinforce our argument that the flyby is an influential interaction type in addition to the merger.

\item
The ratio of the flyby fraction to the merger fraction also shows dependencies on the halo mass, environmental parameter, and redshift, led mainly by the flyby fraction. 
The flyby contributes to the pair fraction from 0.4\,\% to 62\,\% (on average, $\sim$\,50\,\%).
The flyby gives a greater contribution to the pair fraction (a) for less massive halos, (b) in the higher density environment, and (c) at lower redshift.

\end{enumerate}

In conclusion, our findings provide a new perspective that the flyby substantially outnumbers the merger for one-fourth of the parameter space toward $z$\,=\,0.
The flyby's fractional contribution to the dynamical evolution of galaxies at the present epoch should be stronger than ever.
The importance of flybys is also highlighted by their contribution to multiple interactions.
Although the flyby is difficult to identify observationally, disk substructures (e.g., warps and tidal tails) and enhancement of the star formation rate can be consequences of the flyby events \citep[e.g.,][]{Kim17}.
In upcoming papers in this series, we will further assess quantitatively the dynamical (e.g., the angular momentum) and hydrodynamic (e.g., star formation rate) effects of the flyby as well as the merger.

\acknowledgments
We thank the referee for helpful comments and suggestions to improve the quality of the paper.
S.-J.Y. acknowledges support from the Mid-career Researcher Program (No. 2019R1A2C3006242) and the SRC Program (the Center for Galaxy Evolution Research; No. 2017R1A5A1070354) through the National Research Foundation of Korea. 

\appendix

\begin{figure*}[htb!]
\centering
\includegraphics[width=0.9\textwidth]{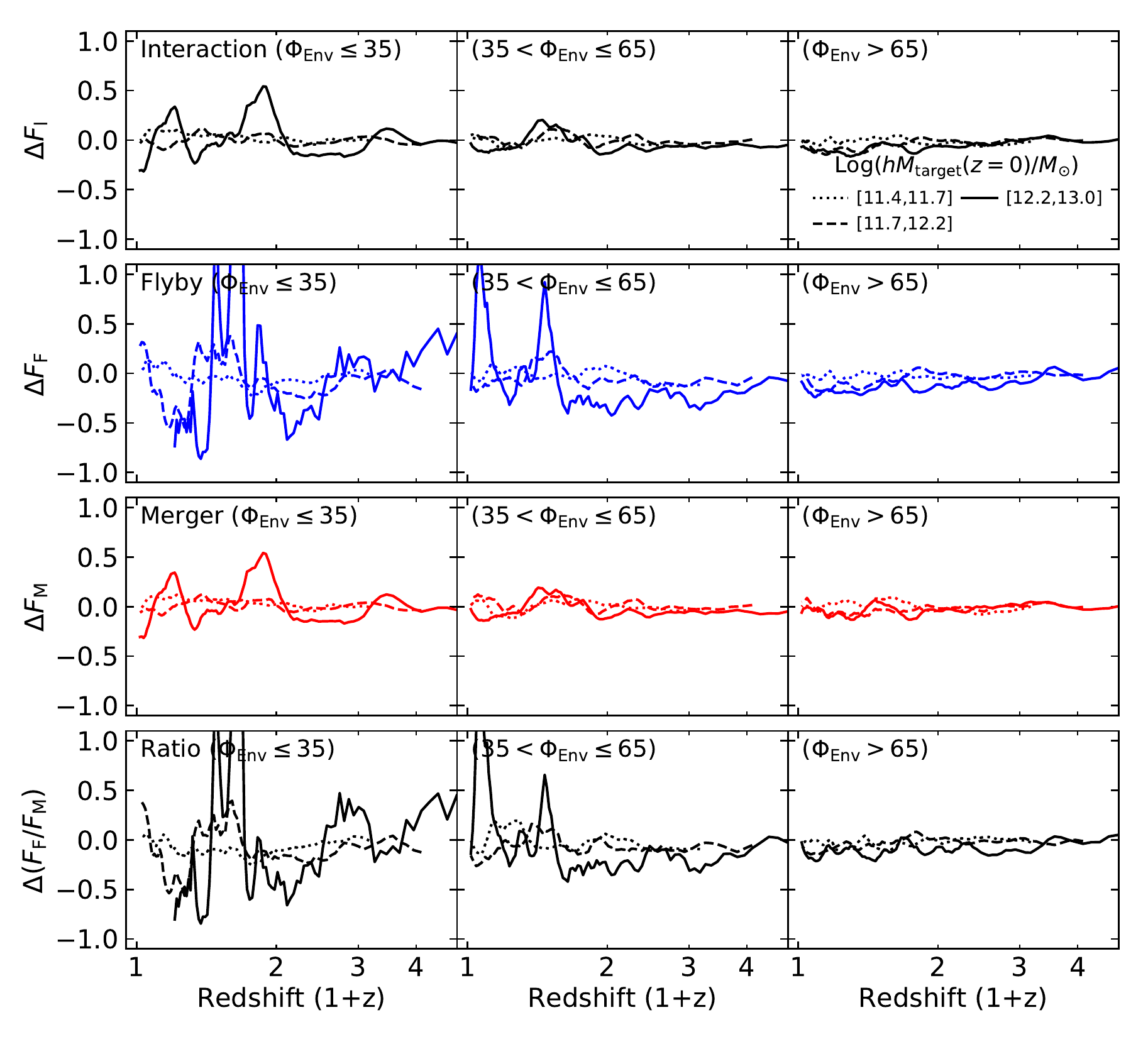}
\caption{Same as Figure \ref{fig:redevol}, but for the difference in the interaction fraction ($\Delta F_{\rm I}$; top row), flyby fraction ($\Delta F_{\rm F}$; second row), merger fraction ($\Delta F_{\rm M}$; third row), and ratio between two fractions ($\Delta(F_{\rm F}/F_{\rm M})$; bottom row).
$\Delta F$ is the difference between the value obtained from 10 GOTPM512 simulations and the value obtained from one GOTPM1024 simulation.
\label{fig:diffrac}}
\end{figure*}

\begin{figure*}[ht!]
\includegraphics[width=0.95\textwidth]{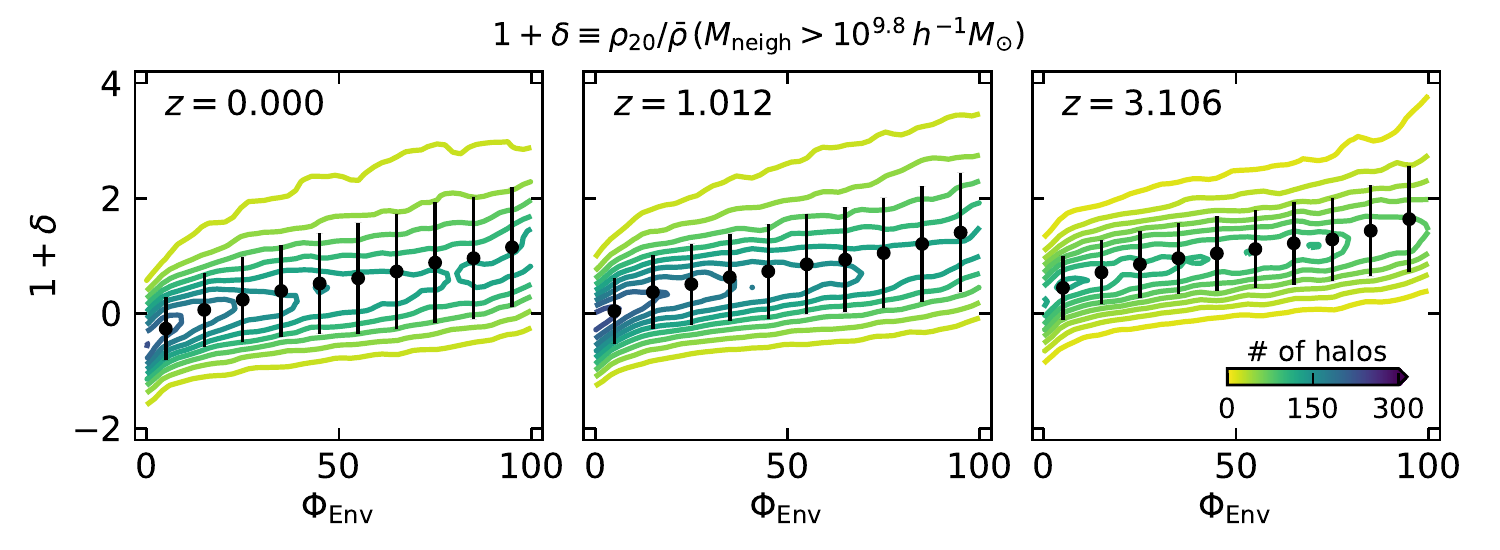}
\centering
\caption{Comparison between our environmental parameter, $\Phi_{\rm Env}$, and another popular definition of the environment, $1+\delta$. 
The latter definition is the ratio between the ambient background density enclosing 20 close neighbors and the mean density. 
Colored contours indicate the number of target halos. 
Black large dots and error bars are the mean values and $1\sigma$ scatters of $1+\delta$ in 10 bins with the same width of $\Phi_{\rm Env}$.}
\label{fig:envcomp}
\end{figure*}

\section{Effect of simulation box size \label{sec:AA}}

In this section, we examine to what extent the simulation box size affects the interaction fractions. 
The cosmic variance and the finite box size effect are inevitable in cosmological simulations.
As mentioned in Section 2, the cosmic variance is reduced by using total 11 simulations (10 GOTPM512 and one GOTPM1024) by a factor of four.
However, the simulation box size effect still remains and is more prominent in GOTPM512 simulations than in the GOTPM1024 simulation because the larger simulation box better reproduces the universe without truncation of the large-scale power \citep{Bag05,Pow06}.

To quantify the simulation box size effect, we compare between the results of 10 GOTPM512 simulations and those of one GOTPM1024 simulation.
Figure \ref{fig:diffrac} shows the differences between GOTPM512 and GOTPM1024 simulations in three values ($F_{\rm I}$, $F_{\rm F}$, $F_{\rm M}$, and $F_{\rm F}/F_{\rm M}$). The difference is calculated by
\begin{equation}
\Delta F = \frac{F({\rm GOTPM512}) - F({\rm GOTPM1024})}{F({\rm GOTPM1024})}\,, 
\end{equation}
where $F({\rm GOTPM512})$ is a value obtained from 10 GOTPM512 simulations, and $F({\rm GOTPM1024})$ is the value obtained from one GOTPM1024 simulation. 
The differences ($\Delta F_{\rm I}$, $\Delta F_{\rm F}$, $\Delta F_{\rm M}$, and $\Delta (F_{\rm F}/F_{\rm M})$) vary with the halo masses, environments, and redshifts.
In terms of the halo mass, the high-mass halos (solid lines) show larger differences than the low-mass halos (dotted lines). 
The standard deviation ($\sigma(F)$) of the difference over cosmic time for the high-mass halos is on average two times greater than that for the low-mass halos.
In terms of the environment, the differences in the low-density environment (left columns) are larger than those in the high-density environment (right columns). 
Also, $\sigma(F)$ in the low-density environment is largely enhanced by a factor of two.
In terms of redshift, the differences decrease with redshift.

The fluctuations of four differences ($\Delta F_{\rm I}$, $\Delta F_{\rm F}$, $\Delta F_{\rm M}$, and $\Delta (F_{\rm F}/F_{\rm M})$) are likely associated with the number of halos.
As the number of halos decreases with increasing halo masses and decreasing redshifts, the fluctuation increases accordingly. 
However, the larger fluctuation for the lower density environment ($\sigma(F_{\rm I})$\,$\sim$\,0.20, $\sigma(F_{\rm F})$\,$\sim$\,0.87, $\sigma(F_{\rm M})$\,$\sim$\,0.20, $\sigma(F_{\rm F}/F_{\rm M})$\,$\sim$\,0.85) cannot be explained solely by changes in the number of halos. 
It seems that GOTPM512 simulations do not reproduce satisfactorily the low-density environment rather than the high-density environment.
Although GOTPM512 simulations do not include the biggest clusters ($M_{\rm halo} \sim 10^{15} \Msun$), $\sigma(F)$ in the high-density (intermediate-density) environment is $\sim$\,0.05 (0.09) on average, and the high-density and intermediate-density environments seem well described.
This implies that missing the larger-scale power in the simulation affects the interaction fraction more severely in the low-density environment than in the high-density environment.
Therefore, the effect of small simulation box size does not change significantly the interaction fractions in dense environments.
Just in passing, we note that a larger simulation box size is necessary for a statistical study of our Galaxy and Andromeda in a cosmological context.

\section{Dependence on the Definition of Environment \label{sec:AB}}

We have used the environmental parameter ($\Phi_{\rm Env}$) defined by Equations \eqref{eq:rank} and \eqref{eq:cmass}. 
In Figure \ref{fig:envcomp}, our parameter is compared with another common environmental definition ($1+\delta$), which is a relative contrast to the mean density \citep[e.g.,][]{Par07,LHu15}. The value of $1+\delta$ is defined by
\begin{equation} \label{envdelta}
1+\delta \equiv \frac{\rho_{20}}{\bar{\rho}}\,\,,
\end{equation}
where $\rho_{20}$ is the ambient background density within an enclosed region by up to the 20th-nearest-neighbor, and $\bar{\rho}$ is the mean density.
The two environmental parameters are positively correlated with each other at all redshifts.
However, the $1\sigma$ scatter is quite large because the two definitions have different intentions; ours is for the large-scale density, but $\rho_{20}$ represents the local density. 
The effective radii of the 20th-nearest-neighbor are considerably smaller than $5\Mpc$ used in our definition \citep{Par07}, implying that $\rho_{20}$ produces the local density in a smaller region than $5\Mpc$ of $\Phi_{\rm Env}$.
Despite the different intentions, the two definitions alike describe well the cosmic web types (see our Figure \ref{fig:snap} and Figure A1 of \citealt{LHu15}), and thus the comparison between our results and those of \citet{LHu15} seems reasonable.

\begin{figure*}[ht!]
\includegraphics[width=0.95\textwidth]{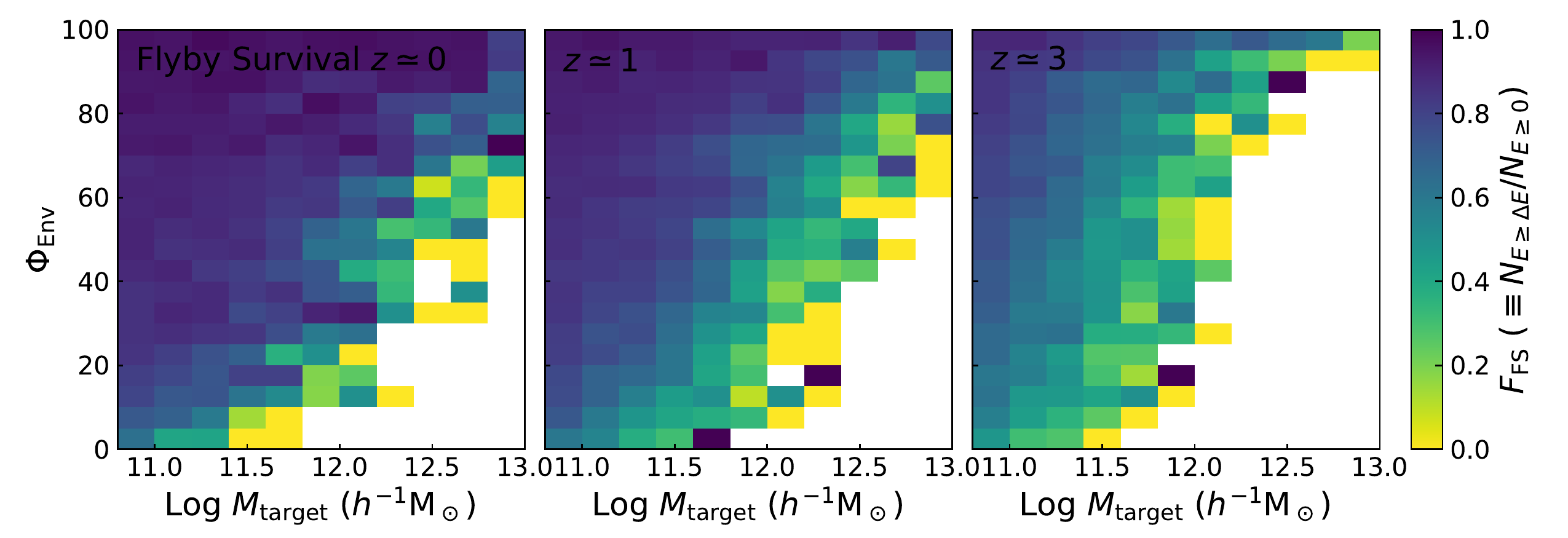}
\centering
\caption{Same as Figure \ref{fig:massenv}, but for the FS fraction ($F_{\rm FS}\equiv N_{E\geq\Delta E}/N_{E\geq0}$), which is the ratio between the number of target halos ($N_{E\geq\Delta E}$) with interacting neighbors with total energy greater than the capture criterion and the number of target halos ($N_{E\geq0}$) with interacting neighbors with a positive energy. Bins with the number of target halos smaller than 50 or without neighbor with positive energy ($N_{E\geq0}=0$) are omitted and left as empty blocks.
\label{fig:dfmenv}}
\end{figure*}

\begin{figure*}[ht!]
\includegraphics[width=0.95\textwidth]{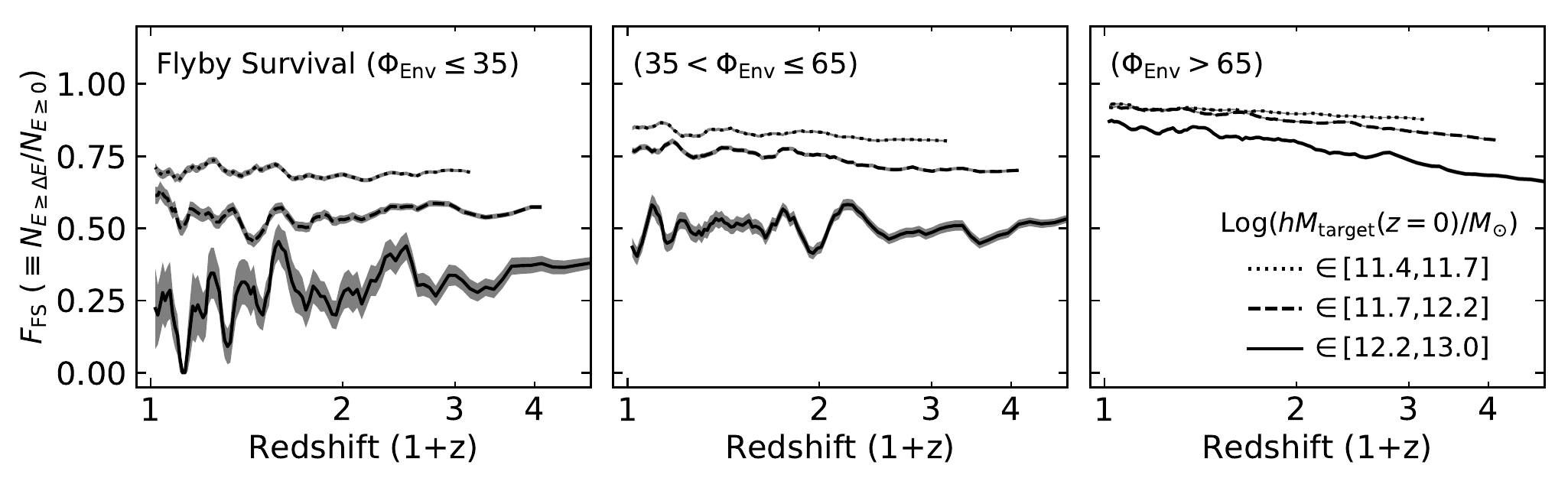}
\centering
\caption{Same as Figure \ref{fig:redevol}, but for the FS fraction ($F_{\rm FS}\equiv N_{E\geq\Delta E}/N_{E\geq0}$). 
\label{fig:dfred}}
\end{figure*}

\section{Effect of Dynamical Friction \label{sec:AC}}

We have applied the capture criterion to the total energy to determine whether a neighbor is flybying or merging.
The dynamical friction can cause some pairs with a positive total energy to be merged.
We define the flyby survival (FS) fraction by
\begin{equation} \label{fs}
F_{\rm FS} \equiv \frac{N_{E\geq\Delta E}}{N_{E\geq0}}\,,
\end{equation}
where $N_{E\geq\Delta E}$ is the number of target halos with neighbors with a total energy greater than the capture criterion, and $N_{E\geq0}$ is the number of target halos with neighbors with a positive total energy.

Figure \ref{fig:dfmenv} shows the FS fraction in the mass--environment plane at three different redshifts.
Regardless of redshift, the FS fraction is higher for the lower mass halos and in the higher density environments than their counterparts, respectively.
In other words, for more massive halos and lower large-scale density, the pair systems more easily lose their energy by dynamical friction.
As mentioned in Section 5.1, owing to the deeper potential of more massive halos and the lower relative velocity in the lower density environment, it is difficult for the total energy to exceed the capture criterion.
Thus, the total energy becomes negative finally.
In contrast, the less massive halo in the denser environment has a shallow potential and a high peculiar velocity, and thus when it interacts with a comparable-mass neighbor, the total energy may easily surpass the capture criterion. 
On the other hand, the FS fraction tends to increase as $z=0$ is approached. The mean values of $F_{\rm FS}$ are 0.89, 0.85, and 0.75 at $z\simeq0$, 1, and 3, respectively. 
This is consistent with \citet{Sin12}, who found that $\sim$\,20\,\% of flybys at $z>1$ are finally merged, suggesting that 80\,\% of the flybys sustain their state.

In Figure \ref{fig:dfred}, we further analyze the redshift evolution of the FS fraction. 
In the low-density environment, the FS fractions are almost constant for the low- and intermediate-mass halos ($F_{\rm FS}\sim$ 0.69 and 0.55) but increase with redshift for the high-mass halos.
In the intermediate-density environment, the FS fractions for the low- and intermediate-mass halos slightly decrease with redshift. 
For the high-mass halos, $F_{\rm FS}$ is roughly 0.51 at all redshifts.
In the high-density environment, the decreasing trend is evident for all mass ranges; on average, $F_{\rm FS}(z=0)\sim0.91$ and $F_{\rm FS}(z=2)\sim0.82$.
This is likely associated with the redshift evolution of median relative velocities (see Figure \ref{fig:medrv}). 
A dramatic drop in the relative velocity may lead to the reduction of $F_{\rm FS}$.

\bibliography{ref}

\clearpage

\end{document}